\documentclass[a4paper,fleqn,usenatbib]{mnras}

\usepackage{newtxtext,newtxmath}
\usepackage[T1]{fontenc}
\usepackage{ae,aecompl}

\usepackage{graphicx}	
\usepackage{amsmath}	
\usepackage{amssymb}
\usepackage{amsfonts}
\usepackage{dcolumn}
\usepackage{color}
\usepackage{latexsym}
\usepackage{placeins}
\usepackage{epsfig}
\usepackage{subfigure, rotating, hyperref, bm, array}
\usepackage{cancel}
\usepackage{soul}
\usepackage[normalem]{ulem}
\usepackage{tikz}
\usetikzlibrary{arrows, shapes}

\newcommand{\stkout}[1]{\ifmmode\text{\sout{\ensuremath{#1}}}\else\sout{#1}\fi}

\title[Shadows and images of naked singularities]{Shadows of spherically symmetric black holes and naked singularities}

\author[R. Shaikh et al.]{
Rajibul Shaikh,$^{1}$\thanks{E-mail: rajibul.shaikh@tifr.res.in}
Prashant Kocherlakota,$^{1}$\thanks{E-mail: k.prashant@tifr.res.in}
Ramesh Narayan$^{2}$\thanks{E-mail: rnarayan@cfa.harvard.edu}
and Pankaj S Joshi$^{1,3}$\thanks{E-mail: psj@tifr.res.in}
\\
$^1$Tata Institute of Fundamental Research, Homi Bhabha Road,
Colaba, Mumbai 400005, India\\ $^2$Harvard-Smithsonian Center for
Astrophysics, 60 Garden Street, Cambridge, MA 02138, USA\\ $^3$International Center for Cosmology,
Charusat University, Anand 388421, Gujarat, India}

\date{Accepted XXX. Received YYY; in original form ZZZ}

\pubyear{2018}

\begin{document}
\label{firstpage}
\pagerange{\pageref{firstpage}--\pageref{lastpage}}
\maketitle

\begin{abstract}
We compare shadows cast by Schwarzschild black holes with those produced by two classes of naked singularities that result from gravitational collapse of spherically symmetric matter.  The latter models consist of an interior naked singularity spacetime restricted to radii $r\leq R_b$, matched to Schwarzschild spacetime outside the boundary radius $R_b$. While a black hole always has a photon sphere and always casts a shadow, we find that the naked singularity models have photon spheres only if a certain parameter $M_0$ that characterizes these models satisfies $M_0\geq 2/3$, or equivalently, if $R_b\leq 3M$, where $M$ is the total mass of the object. Such models do produce shadows. However, models with $M_0<2/3$ (or $R_b>3M$) have no photon sphere and do not produce a shadow.  Instead, they produce an interesting ``full-moon'' image. These results imply that the presence of a shadow does not by itself prove that a compact object is necessarily a black hole. The object could be a naked singularity with $M_0\geq 2/3$, and we will need other observational clues to distinguish the two possibilities.  On the other hand, the presence of a full-moon image would certainly rule out a black hole and might suggest a naked singularity with $M_0<2/3$.  It would be worthwhile to generalize the present study, which is restricted to spherically symmetric models, to rotating black holes and naked singularities.
\end{abstract}



\section{Introduction}
Currently, there is compelling evidence for the presence of compact regions in the Universe with very large mass, and this is interpreted as strong indirect evidence for the existence of black holes. The compact object Sagittarius A$^*$ (Sgr A$^*$) at our Galactic Center, with a mass of $4\times10^6M_\odot$, is the best example of such an object \citep{schodel+02, ghez+08}, and there is considerable evidence for similar objects of even greater mass at the centers of many other galaxies \citep{kormendy_ho13}.  However, direct evidence for the presence of a black hole requires actual detection of the event horizon, the surface that encloses the compact interior of the black hole, and from where no material particles or light rays can escape. A number of tests have been proposed to confirm the presence of event horizons in black hole candidates \citep{narayan+08, broderick+09, broderick+15}. The evidence is strong but, of necessity \citep{abramowicz+02}, not conclusive.

With the purpose of strengthening the evidence for the presence of a black hole in Sgr A$^*$, as well as in the nucleus of the nearby galaxy M87, the event horizon telescope (EHT, \citealt{EHT1, EHT2}), an Earth-spanning millimeter-wave interferometer, is being constructed and has begun collecting data.  While nothing escapes from the interior of a black hole, the exterior spacetime has a photon sphere which is predicted to create a characteristic shadow-like image of the radiation emitted by an accretion flow around the black hole \citep{bardeen+72, luminet79, falcke+00}. The goal of the EHT is to verify the presence of this shadow at mm wavelengths in the image of Sgr A$^*$. This would add considerably to the evidence that Sgr A$^*$ is a black hole.

Typically, compact black hole-like objects in the Universe are created by gravitational collapse of matter, e.g., from the collapse of massive stars at the end of their life-cycles. Having exhausted their internal nuclear fuel, these stars end up as compact stellar-mass objects. Alternatively, clustering of matter in the central region of a galaxy may generate a massive compact object, which may then grow
further in mass by accretion.

It is worth noting here that a major and frontier unresolved problem in gravitation physics, at the very foundation of black hole physics and its astrophysical applications, is showing that all physically
reasonable gravitational collapses end up producing a black hole only. In that case, the spacetime singularity resulting from collapse (a necessary implication of general relativity) will always be hidden within an event horizon, allowing no signals to be seen from the vicinity of the spacetime singularity. The hypothesis that this is necessarily so is called the cosmic censorship conjecture \citep{Penrose_1969}. When the singularity of collapse is not covered by an event horizon, it is called a naked singularity, which is in principle visible to faraway observers in the Universe. The exciting prospect in the latter case is the possibility of having observable signatures from ultra-strong gravity regions near the singularity. Despite five decades of serious efforts, cosmic censorship remains an unproven conjecture, to the extent that we do not even have a definite mathematical formulation of the conjecture.      

On the other hand, many studies of gravitational collapse have been carried out and it turns out that the end state of collapse is not necessarily always a black hole.  This important key issue was studied in considerable detail in past decades \citep{Christodoulou84, Ori_Piran87, Choptuik93} and the formation of event horizons, trapped surfaces and apparent horizons has been investigated in different scenarios \citep{JM_2011, J_2007}. While the general theory of relativity necessarily predicts the occurrence of a spacetime singularity as a result of collapse 
for a wide range of initial conditions, it turns out that the formation or otherwise of an event horizon, as well as the actual epoch of horizon formation, are governed by the specific regular initial conditions from which the collapse evolves. Therefore, the final state of continual collapse turns out to be either a black hole or a naked singularity, depending on the initial conditions and the allowed evolutions by the Einstein equations, in many physically reasonable collapse scenarios.  

Recently, we investigated physically reasonable gravitational collapse scenarios that end up as naked singularities \citep{JMN,JMN_pressure}, and computed various properties of these spacetimes, such as the nature of stable circular orbits and the spectra of accretion disks. We found that spectra, in particular, may be helpful to discriminate between black holes and naked singularities.

The study of shadows and images of compact objects has been a subject of great interest. The optical appearance of a star collapsing through its gravitational radius was first studied by \citet{Ames_Thorne+68}%
, and of a star orbiting an extreme Kerr black hole was discussed by \citet{Cunningham_Bardeen+73}. %
Many authors \citep{Synge+66, Zakharov+14, Takahashi+04, Hioki_Maeda+09, Takahashi+05, deVries+00, Young+76, Yumoto+12, Shipley+16} have studied the characteristics of shadows cast by various black holes. Structures of shadows and images of black holes have been discussed in the context of determining their spins and masses, and in testing general relativity \citep{Huang+07, Kamruddin_Dexter+13, Johannsen_Psaltis+10}. Implications of black hole shadows on the distribution of dark matter have been explored by \citet{Lacroix_Silk+13}. Also, \citet{Schneider_Perlick+18} have calculated the time-dependence of the angular radius of the shadow in the course of formation of a black hole from gravitational collapse.

One of the early explorations of images and shadows cast by naked singularities was by \citet{Nakao+03}. They studied how the central naked singularity that formed during the collapse of a self-similar dust cloud, was observed by distant observers. By investigating radial and non-radial null geodesics emanating from the singularity, they were able to show that the angular diameter of the image is time dependent; it grows monotonically and approaches the value $3\sqrt{3}M/R_o$ for an observer at $R= R_o \gg M$. The asymptotic value of the angular diameter comes from the geometry of the exterior Schwarzschild region. %
Later,
\citet{Kong+14} studied the radiation emitted by collapsing spherically symmetric dust clouds evolving from different initial data, leading to both black holes and naked singularities as end states. They found that within their simplified model, both these objects had very similar observational features and that it was difficult to differentiate between them based on their light curves. %
\citet{Ortiz+15a} addressed how the redshift of photons travelling from past to future null infinity through a ball of collapsing dust could provide an observational signature capable of differentiating between the formation of a globally naked singularity from the formation of an event horizon. \citet{Ortiz+15b} also pointed out that, although at late times the image of the source perceived by the observer looks the same in both cases, the dynamical formation of the shadow and the redshift images has distinct features and time scales. %
Effects of gravitational lensing around naked singularities have also been investigated \citep{Virbhadra_Keeton+08, Virbhadra_Ellis+02}. Shadows cast by the overspinning Kerr geometry with its central singularity excised was considered by \citet{Bambi_Freese+09}.

\citet{Broderick-Narayan} studied the shadows and images cast by a compact object with a thermally emitting surface and compared them with those of a black hole. They showed that, in some cases, the images could be nearly identical. %
\citet{Saida+16} examined the geometry outside compact objects modeled using a static spherical polytropic perfect fluid and found that they admit no photon spheres and therefore cast no shadows, allowing them to be distinguishable from black holes. %
\citet{SST+14} explored the subject of whether there are supercompact objects, that are not black holes, which possess unstable circular orbits of photons, and how one can distinguish them from black holes based on their shadows. Considering the spherical thin-shell model of a gravastar, \citet{Visser_Wiltshire+04} found that unstable circular orbits of photons can appear around the gravastar, and that one could tell the difference between a black hole and a gravastar with high-resolution very-long-baseline-interferometry observations in the near future. Shadows cast by horizonless exotic compact objects such as wormholes have also been a subject of great interest \citep{intensity_formula2, ohgami_2015, ohgami_2016, Shaikh_2018}.

The purpose of the present work is to examine shadows and images of the naked singularity models of \citet{JMN,JMN_pressure}, and to compare them with the images we expect from black holes. The goal is to check whether the images corresponding to the two kinds of model are clearly distinguishable. Our interesting conclusion is that, while black holes always cast a shadow, naked singularities may or may not, depending on the specific structure of the singularity. Therefore, while black holes imply shadows, the converse is not true. A shadow could be produced by certain naked singularities as well.

The plan of the paper is as follows. In \S\ref{sec:spacetimes}, we briefly review the collapse models we use here, and outline some of their properties. In \S\ref{sec:shadow_boundary}, we investigate geodesic motion, unstable photon orbits and the resulting shadows. In \S\ref{sec:lensing}, we study gravitational lensing and relativistic images in the various spacetimes. In \S\ref{sec:shadow}, we consider a simple accretion model and compute images, which we use to examine how black holes and naked singularities could be distinguished.  We then repeat the analysis in \S\ref{sec:shadow2} using a more realistic accretion flow model and show that the results are largely unchanged. We conclude in \S\ref{sec:summary} with a summary of the key results.

\section{The black hole and naked singularity spacetimes}
\label{sec:spacetimes}

We compare images and shadows produced by a Schwarzschild black hole with those produced by two different naked singularity spacetimes. The latter two solutions describe the geometry around compact objects formed from gravitational collapse of two different types of fluids.

The first naked singularity solution, which we call JMN-1, is formed from the collapse of matter with zero radial pressure, and is described by the following metric \citep{JMN},
\begin{eqnarray} \label{eq:JMN1}
ds_1^2 &=& -(1-M_0)\left(\frac{r}{R_b}\right)^{M_0/(1-M_0)}
dt^2+\frac{dr^2}{1-M_0} \nonumber \\
&& +r^2\left(d\theta^2+\sin^2\theta
d\phi^2\right), 
\end{eqnarray} 
where the parameter $M_0$ is limited to the range $0 \leq M_0 \leq 4/5$ (the upper limit corresponds to the requirement that the sound speed should not exceed unity).  The matter content of this spacetime has the following energy density $\rho$, radial pressure $p_r$, and tangential pressure $p_\theta$:
\begin{equation} \label{eq:EM_JMN1}
\rho=\frac{M_0}{r^2},\;\;\; p_r=0,\;\;\; p_\theta= \frac{M_0}{4(1-M_0)} \rho=\frac{M_0^2}{4(1-M_0)}\frac{1}{r^2}.
\end{equation}
This fluid has non-zero tangential pressure, but its radial pressure is assumed to vanish.

The second naked singularity solution, which we call JMN-2, is the end state of collapse of a spherical cloud with non-zero radial pressure. It describes, for example, the collapse of a perfect fluid cloud with a locally varying equation of state $k(r) = p/\rho$ (not strictly isothermal) that approaches a constant value in the neighborhood of the center of the cloud.
This spacetime is described by the metric \citep{JMN_pressure},
\begin{eqnarray} \label{eq:JMN2}
ds_{2}^2 &=& -\frac{1}{16\lambda^2(2-\lambda^2)}\times \nonumber \\
&&\left[(1+\lambda)^2\left(\frac{r}{R_b}\right)^{1-\lambda}- (1-\lambda)^2\left(\frac{r}{R_b}\right)^{1+\lambda} \right]^2dt^2 \nonumber \\
&& + (2-\lambda^2)dr^2+r^2\left(d\theta^2+\sin^2\theta d\phi^2\right), \nonumber
\end{eqnarray} 
where $0\leq \lambda < 1$. The expressions for the energy density and pressure can be found in \citet{JMN_pressure}. For easier comparison with the JMN-1 model, we define a parameter $M_0$,
\begin{equation}
M_0 = \frac{1-\lambda^2}{2-\lambda^2},
\label{eq:M0}
\end{equation} 
which represents an alternative way (instead of $\lambda$) of parametrizing JMN-2.

Both JMN-1 and JMN-2 contain a time-like naked singularity at $r=0$ and no trapped surface forms in these spacetime. Both solutions are matched at their outer radius $r=R_b$ to the Schwarzschild geometry,
\begin{equation}
ds_0^2=-\left(1-\frac{2M}{r}\right)
dt^2+\frac{dr^2}{1-\frac{2M}{r}}+r^2\left(d\theta^2+\sin^2\theta
d\phi^2\right).
\end{equation}
In both cases, the total mass $M$ is given by
\begin{equation}
M = \frac{1}{2}M_0 R_b.
\label{eq:MRb}
\end{equation}

The three spacetimes we consider, viz., the two naked singularity spacetimes and the Schwarzschild spacetime, can be written in the general form,
\begin{equation} 
ds_i^2=-f_i(r)dt^2+\frac{dr^2}{g_i(r)}+r^2\left(d\theta^2+\sin^2\theta d\phi^2\right),
\label{eq:general metric}
\end{equation}
where the JMN-1 spacetime corresponds to $i=1$, hence
\begin{equation}
f_1(r) = (1-M_0)\left(\frac{r}{R_b}\right)^{M_0/(1-M_0)},  \nonumber
\end{equation}
\begin{equation}
g_1(r)=(1-M_0),\ \ \ \mbox{(JMN-1),}
\end{equation}
the JMN-2 spacetime corresponds to $i=2$ and has
\begin{equation}
f_2(r) = \frac{1}{16\lambda^2(2-\lambda^2)}\left[(1+\lambda)^2\left(\frac{r}{R_b}\right)^{1-\lambda}- (1-\lambda)^2\left(\frac{r}{R_b}\right)^{1+\lambda} \right]^2, \nonumber 
\end{equation}
\begin{equation}
g_2(r)=\frac{1}{2-\lambda^2},\ \ \ \mbox{(JMN-2),}
\end{equation}
and the Schwarzschild spacetime corresponds to $i=0$ and has
\begin{equation} \label{eq:Sch_gen_form}
f_0(r) = g_0(r) = \left(1-\frac{2M}{r}\right),\ \ \ \mbox{(Schwarzschild).}
\end{equation}

We now briefly discuss a few technical issues that are relevant for our study of images and shadows. First, we address the issues of whether the JMN spacetimes satisfy the Tolman-Oppenheimer-Volkoff equation (TOV, \citealt{Tolman+34, Oppenheimer_Volkoff+39}), which is derived by solving the Einstein equations together with the conservation equations for a general time-independent, spherically symmetric metric (canonically for perfect fluids). Since the gravitational collapse process that leads to the the above JMN spacetimes is studied by solving the Einstein equations together with the conservation equations (See \citealt{JMN, JMN_pressure}), when a time-invariant configuration (equilibrium) is attained, the TOV equation must automatically be satisfied. For the JMN-2 class of spacetimes, the TOV equation at equilibrium is discussed and shown to be satisfied (see Eq. (27) of \citealt{JMN_pressure}). Here, we point out that the JMN-1 spacetime, containing an imperfect fluid, also obeys the anisotropic TOV equation. For the spacetime (\ref{eq:general metric}) supported by the energy-momentum tensor of the form $T^\mu_{\;\nu}=\text{diag}[-\rho,p_r,p_{\theta},p_{\theta}]$, the anisotropic TOV equation is given by (see Eq. (21) of \citealt{Chirenti_Rezzolla+07}),
\begin{equation}
p_r'=-(\rho+p_r)\frac{m(r)+r^3p_r/2}{r(r-2m(r))}+\frac{2}{r}(p_\theta -p_r),
\end{equation}
where we have set $8\pi G=1$ and $c=1$, and
\begin{equation}
g_i(r)=1-\frac{2m(r)}{r}, \quad 2m(r)=\int_0^r \rho r^2 dr.
\end{equation}
For the JMN-1 spacetime, the energy density and pressures are given by (\ref{eq:EM_JMN1}) and $m(r)=\frac{1}{2}M_0 r$. It can immediately be checked that the TOV equation is satisfied by this spacetime.

The JMN-1 spacetime may not be physically realistic since the radial pressure is assumed to vanish (but this assumption simplifies the analysis considerably and allows simple analytical expressions). JMN-2 is more realistic since it is supported by a fluid with isotropic pressure. The matter for the JMN-2 can have an equation of state of the form $p=k\rho$, where $k$ may not necessarily be a constant. Since the pressure must vanish at the surface $r=R_b$ whereas the density may not, we must have $k|_{r=R_b}=0$. Other examples of stable configurations, where the matter has a constant density and isotropic pressure with a variable equation of state, exist in the literature (see \S6.2 of \citealt{Wald} or Eq. (1) of \citealt{Pani_Ferrari+18}).

Next, we show that the JMN spacetimes can be smoothly matched across the $r=R_b$ hypersurface $\Sigma$, i.e., the metric tensor $g_{\mu\nu}$ and the extrinsic curvature $K_{ab}$ are continuous across $\Sigma$. By construction $g_{\mu\nu}$ is continuous across $\Sigma$ (see \ref{eq:general metric}-\ref{eq:Sch_gen_form}). To show that $K_{ab}$ is also continuous, we first note that the coordinates in both the interior and the exterior are $x^\mu=(t,r,\theta,\phi)$, and those on $\Sigma$ are $y^a=(t,\theta,\phi)$. Therefore, as seen from the Schwarzschild exterior, the induced metric on $\Sigma$ is
\begin{equation}
ds^2_\Sigma =-(1-M_0)dt^2+R_b^2(d\theta^2 +\sin^2\theta d\phi^2),
\label{eq:induced_metric}
\end{equation}
where we have used (\ref{eq:MRb}). Also, as seen from the interior JMN spacetimes, the induced metric on $\Sigma$ is the same as (\ref{eq:induced_metric}). This is reminiscent of the fact that the metric tensor $g_{\mu\nu}$ is continuous across $\Sigma$. The non-zero components of the tangent $e^\mu_{\;a}=\partial x^{\mu}/\partial y^a$ on $\Sigma$ are $e^t_{\;t}=1$, $e^\theta_{\;\theta}=1$ and $e^\phi_{\;\phi}=1$. The extrinsic curvature of $\Sigma$ is given by $K_{ab}=e^\mu_{\;a}e^\nu_{\;b}\nabla_{\nu}n_\mu$, where $n^{\mu}$ is a unit normal to $\Sigma$. Now, as seen either from the exterior Schwarzschild or from the interior JMN spacetimes (\ref{eq:general metric}), the unit normal is given by $n^{\mu}=(0,\sqrt{g_i(r)},0,0)$. Therefore, as seen either from the exterior Schwarzschild or from the interior JMN spacetimes, the non-zero components of the extrinsic curvature are given by
\begin{equation}
K^t_{t}=\frac{f_i'(r)}{2f_i(r)}\sqrt{g_i(r)}\Big|_{r=R_b}, \quad K^\theta_{\theta}=K^\phi_{\phi}=\frac{1}{r}\sqrt{g_i(r)}\Big|_{r=R_b}.
\end{equation}
Note that $f_{1,2}(R_b)=f_0(R_b)=(1-M_0)$ and $g_{1,2}(R_b)=g_0(R_b)=(1-M_0)$. Also, it is straighforward to show that $f'_{1,2}(R_b)=f'_0(R_b)=M_0/R_b$, implying that the extrinsic curvature is also continuous across $\Sigma$. Therefore, the JMN spacetimes are smoothly matched to the exterior Schwarzschild spacetime at $r=R_b$.

The JMN spacetimes do not contain any trapping region since $(1-2m(r)/r)=1-M_0>0$ always. However, to show that the singularities are actually naked, we have to show that photons emitted from the singularity, or from its vicinity, reach faraway observers in a finite time. To this end, we calculate both the affine time and the time measured by a faraway static observer. Taking $E=1$ (see \S\ref{sec:shadow_boundary}), as measured by a faraway static observer, the time taken by a radially outgoing photon ($L=0$) to travel from $r$ to the surface is given by
\begin{equation}
\Delta t=t(R_b)-t(r)=\int_r^{R_b} \frac{\dot{t}}{\dot{r}}dr=\int_r^{R_b} \frac{1}{\sqrt{f_i(r)g_i(r)}}dr.
\end{equation}
For JMN-1, this gives
\begin{equation}
\Delta t=t(R_b)-t(r)=\frac{2R_b}{2-3M_0}\left[1-\left(\frac{r}{R_b}\right)^{\frac{2-3M_0}{2(1-M_0)}}\right].
\end{equation}
Note that, for $M_0<2/3$, $\Delta t$ is finite for a photon escaping from the singularity $r=0$ to reach the surface $R_b$. However, for $M_0>2/3$, $\Delta t$ diverges as $r\to 0$, implying that, as measured by a faraway static observer, a photon escaping from the singularity takes infinite time to reach the observer. For JMN-2, we have
\begin{equation}
\Delta t=\frac{2(2-\lambda^2)R_b}{1-\lambda^2}\left[\log\frac{2(1+\lambda)}{\lambda}-\log\frac{(1+\lambda)R_b+(1-\lambda)r}{(1+\lambda)R_b-(1-\lambda)r}\right].
\end{equation}
We see that $\Delta t$ is finite as $r\to 0$, implying that a photon escaping from the singularity always takes finite time to reach a faraway observer. 
We next calculate the affine time $\tau$. From (\ref{eq:radial_eq}), for radial null geodesics, we obtain
\begin{equation}
\Delta\tau=\tau(R_b)-\tau(r)=\int_r^{R_b}\sqrt{\frac{f_i(r)}{g_i(r)}}dr.
\end{equation}
For JMN-1 spacetime, this becomes
\begin{equation}
\Delta\tau=\frac{2R_b(1-M_0)}{2-M_0}\left[1-\left(\frac{r}{R_b}\right)^{\frac{2-M_0}{2(1-M_0)}}\right],
\end{equation}
which is finite always as $r\to 0$, implying that a photon escaping from the JMN-1 singularity always reaches a faraway observer in a finite affine time. For JMN-2, we have
\begin{equation}
\Delta\tau=\frac{R_b(1+\lambda)^2}{4\lambda(2-\lambda)}\left[1-\left(\frac{r}{R_b}\right)^{2-\lambda}\right]-\frac{R_b(1-\lambda)^2}{4\lambda(2+\lambda)}\left[1-\left(\frac{r}{R_b}\right)^{2+\lambda}\right],
\end{equation}
which is again finite as $r\to 0$, implying that a photon escaping from the JMN-2 singularity always reaches a faraway observer in a finite affine time. 

Thus, except for the JMN-1 spacetime with $M_0>2/3$, both the coordinate time $t$ and the affine time $\tau$ are finite for a photon escaping from the singularities and reaching a faraway observer. In the case of JMN-1 spacetime with $M_0>2/3$ however, even though the coordinate time $t$ is infinite, the affine time $\tau$ is finite for photons emitted from an infinitesimally close vicinity of the singularity ($r\simeq 0$). In this case, the behaviour of null geodesics escaping from the singularity is very similar to that of null geodesics escaping from the event horizon of a Schwarzschild black hole. However, photons emitted from a finitely close vicinity ($r\sim 0$) of the singularity will take large but finite coordinate times $t$ to reach a faraway observer.


Finally, it is worth emphasizing that the spacetimes we study here are merely toy models that we use to explore potential observational signatures of naked singularities. At this stage of our study, we do not view astrophysical realism as an important requirement. The main virtue of these models is that, they are not merely exact solutions of the steady state (time-independent) Einstein equations (of which there are many), but we have shown that these solutions develop via time evolution from regular initial conditions \citep{JMN, JMN_pressure}. The latter is a rather stringent requirement.
As is well-known, the Schwarzschild black hole metric forms via evolution from non-singular initial conditions, e.g., the famous Oppenheimer-Snyder model. However, none of the other black hole solutions (Reissner-Nordstrom, Kerr) has been shown to form in their entirety, i.e., both outside and inside the horizon, from regular initial conditions. Similarly, we do not believe that many of the naked singularity models in the literature, e.g., Kerr with $a>M$, Reissner-N{\"o}rdstrom with $Q > M$ and the Janis-Newman-Winicour naked singularity have been shown to form from physically well-behaved initial conditions. In contrast, our naked singularity solutions do form from perfectly regular initial conditions, as we have demonstrated in our previous work. We view this as a major advantage of these models, and in this sense we consider our models to be ``physically realistic".

\section{Shadows of JMN naked singularities and Schwarzschild black hole}
\label{sec:shadow_boundary}

The shadow structures for the different spacetimes are determined by the properties of null geodesics in these spacetimes.  We therefore begin with a discussion on this topic.  We consider an extended source of radiation on the far side of the compact object. Photons from the source traverse the spacetime of the black hole or naked singularity, get deflected, and reach the observer. As viewed by the observer, we are interested in those directions for which no (or very little) radiation is received.  The union of these directions constitutes the shadow of the gravitating object.

\subsection{Geodesic motion and unstable photon orbits}
The Lagrangian describing the motion of a photon in the spacetime geometry \eqref{eq:general metric} is given by
\begin{equation}
2\mathcal{L}=-f_i(r)\dot{t}^2+\frac{\dot{r}^2}{g_i(r)}+r^2\dot{\theta}^2 + r^2\sin^2\theta \dot{\phi}^2,
\end{equation}
where a dot represents a derivative with respect to the affine parameter. Since the Lagrangian is independent of $t$ and $\phi$, we have two constants of motion:
\begin{equation}
p_t=\frac{\partial\mathcal{L}}{\partial\dot{t}}=-f_i(r)\dot{t}=-E,
\end{equation}
\begin{equation}
p_\phi=\frac{\partial\mathcal{L}}{\partial\dot{\phi}}=r^2\sin^2\theta \dot{\phi}=L,
\end{equation}
where $E$ and $L$ are, respectively, the energy and angular momentum of the photon.  Using the null geodesic condition $g_{\mu\nu}\dot{x}^\mu\dot{x}^\nu=0$, we obtain
\begin{equation}
\frac{1}{g_i}\dot{r}^2+r^2\dot{\theta}^2=\frac{r^2\sin^2\theta E^2-f_i L^2}{f_i r^2\sin^2\theta}.
\label{eq:r_theta}
\end{equation}
Since our spacetimes are spherically symmetric, the shadows and images will be circularly symmetric in the observer sky. Thus the intensity will be a function only of the impact parameter $b=L/E$ with respect to the center of the spacetime, and will be independent of the azimuthal angle $\theta$.  Therefore, we can simply choose $\theta=\pi/2$, $\dot{\theta}=0$, and obtain all our results for this
case. The same results can then be applied to all $\theta$.  Setting $\theta=\pi/2$ and $\dot{\theta}=0$, we obtain
\begin{equation} \label{eq:radial_eq}
\frac{f_i}{g_i}\dot{r}^2+V_{eff}=0, \hspace{0.3cm}
V_{eff}={L^2}\frac{f_i(r)}{r^2}-E^2.
\end{equation}
The impact parameter $b$ can be related to the turning point $r_{tp}$ of a photon, where $\dot{r}=0$ and $V_{eff}(r_{tp})=0$:
\begin{equation}
b=\frac{r_{tp}}{\sqrt{f_i(r_{tp})}}.
\label{eq:impact_turning}
\end{equation}
This expression will be useful in our subsequent analysis. Circular photon orbits satisfy $V_{eff}=0$ and $dV_{eff}/dr=0$, and we have,
\begin{equation}
x f_{i,x}-2 f_i=0,
\label{eq:PhSph_zero_spin1}
\end{equation}
\begin{equation}
{\frac{b^2}{R_b^2}}=\frac{x^2}{f_i},
\label{eq:PhSph_zero_spin2}
\end{equation}
where $x=r/R_b$, and $f_{i,x}$ represents differentiation of $f_i$ with respect to $x$. Note that the matching surface between the interior naked singularity spacetime and the exterior Schwarzschild
spaceime is now at $x=x_b=1$.  The photon sphere comprises of circular unstable photon orbits, i.e., circular orbits that satisfy additionally $d^2V_{eff}/dr^2<0$.

Equation \eqref{eq:PhSph_zero_spin1} does not have any non-trivial solution for the two interior JMN spacetimes we are considering. Therefore, there is no photon sphere for either interior JMN spacetime. However, the JMN spacetimes are matched on the exterior to the Schwarzschild geometry, and the latter spacetime does have unstable photon orbits on a photon sphere located at
\begin{equation}
r_{ph} = 3M, \qquad x_{ph} \equiv \frac{r_{ph}}{R_b} = \frac{3}{2}M_0. 
\end{equation}
Therefore, the existence or not of a photon sphere in the naked singularity models depends on the relative sizes of $x_{ph}$ and the matching radius $x_b$.  A photon sphere exists whenever the following conditions, which are all equivalent, are satisfied
\begin{equation}
R_b \leq 3M, \qquad x_{ph} \geq x_b, \qquad M_0 \geq \frac{2}{3}.
\end{equation}
There is no photon sphere when $M_0 < 2/3$, or equivalently, when $R_b > 3M$.

The JMN-1 spacetime satisfies reasonable physical conditions (e.g., sound speed less than unity) for the parameter range $0<M_0<4/5$. The subset of these models with $2/3 \leq M_0\ < 4/5$ have photon spheres, while the rest do not. The JMN-2 spacetime is parametrised by $0 \leq \lambda = \sqrt{(1-2M_0)/(1-M_0)}\leq 1$, which means that the allowed range of $M_0$ is $0\leq M_0\leq 1/2$. Thus, JMN-2 is devoid of a photon sphere for the entire allowed range of parameter values. It
should be noted that for the cases for which photon spheres exist, they are always located in the exterior Schwarzschild geometry. Setting $x = x_{ph} = 3M_0/2$ in Eq. \eqref{eq:PhSph_zero_spin2}, we obtain
\begin{equation}
{b_{ph}^2}=\frac{27}{4}M_0^2R_b^2=27M^2,
\label{eq:xi_eta}
\end{equation}
which is the same equation as that obtained for the Schwarzschild black hole. The only difference in the case of the JMN spacetimes is that we have the additional requirement, $M_0\geq2/3$ (or $R_b\leq
3M$), in order to have a photon sphere. In the above discussion, {$b_{ph}$} is the critical impact parameter of a photon on an unstable photon orbit.

\subsection{Shadows}
The unstable photon orbits constitute the photon sphere, and they define the boundary of the shadow cast by a compact object. Photons from a distant source with impact parameter {$b$} larger than the
critical impact parameter $b_{ph}$, i.e.,
\begin{equation}
{b^2 > 27 M^2,}
\end{equation}
remain outside the photon sphere and reach the observer. However, photons with impact parameters smaller than the critical impact parameter are captured within the photon sphere and do not reach the
observer, thereby creating dark spots in the observer's sky. The union of these dark spots constitutes the shadow. Therefore, the apparent shape of the shadow projected in the observer's sky is a circular disk whose radius is given by the critical impact parameter $b_{ph}=3\sqrt{3}M$.

\begin{figure}
\centering
\subfigure[~JMN-1 naked singularity]{\includegraphics[scale=0.67]{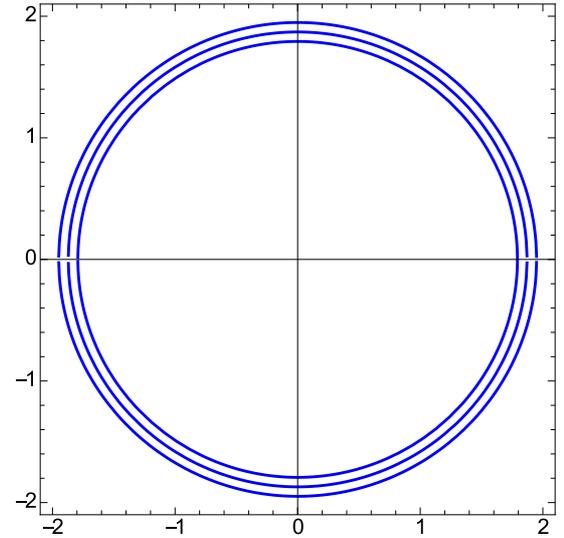}}\hspace{0.1cm}
\subfigure[~Schwarzschild black hole]{\includegraphics[scale=0.67]{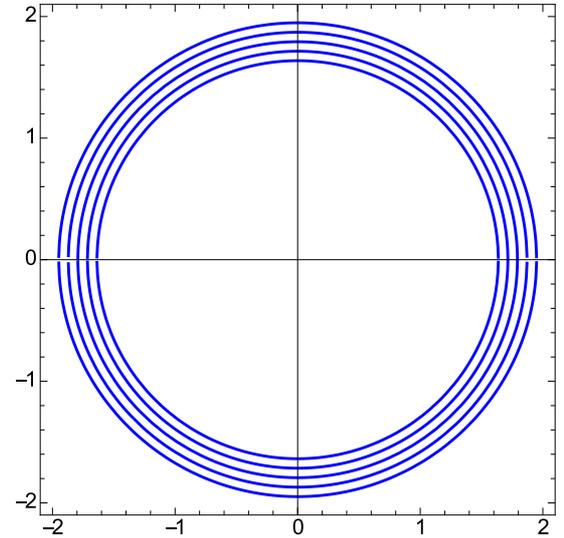}}
\caption{Shadows cast by (a) the JMN-1 naked singularity, 
matched to the exterior Schwarzschild spacetime at $x=x_b=1$, for $M_0=0.75,0.72,0.69, 0.66,0.63$ (from outer to inner), in the units of $R_b$, and (b) the Schwarzschild black hole with mass $M$ equal to the same set of values (from outer to inner). For $M_0<2/3$, the JMN-1 naked singularity does not cast any shadow. 
}
\label{fig:shadowA}
\end{figure}

Figure \ref{fig:shadowA} shows the shapes of shadows cast by the JMN-1 naked singularity and the Schwarzschild black hole. The circles represent the outer boundaries of the shadows. In the case of the
black hole, shadows exist for all $M$. However, in the case of the JMN-1 naked singularity, depending on the value of $M_0$, a shadow may either form ($M_0 \geq 2/3$), or not form ($M_0<2/3$). As we noted earlier, in the case of the JMN-2 naked singularity there is no photon sphere, and therefore this spacetime does not cast any shadow.

\section{Gravitational lensing and relativistic images}
\label{sec:lensing}
Since the shadows and images are the result of strong gravitational lensing, we now study lensing by the JMN naked singularities. From  \S\ref{sec:shadow_boundary}, we obtain
\begin{equation}
\frac{d\phi}{dr}=\frac{1}{r^2\sqrt{g_i(r)}}\frac{1}{\sqrt{\frac{1}{b^2f_i(r)}-\frac{1}{r^2}}},
\end{equation}
where $b$ is the impact parameter. Defining $u=R_b/r$, we obtain the deflection angle
\begin{equation}
\alpha=2\int_0^{u_{tp}}\frac{1}{\sqrt{g_i}}\frac{1}{\sqrt{\frac{1}{\bar{b}^2f_i(u)}-u^2}}du-\pi,
\label{eq:deflection_junction1}
\end{equation}
where 
\begin{equation}
\bar{b} = \frac{b}{R_b}, \qquad u_{tp} = \frac{R_b}{r_{tp}},
\end{equation}
and $r_{tp}$ is the turning point given by $dr/d\phi=0$.  For the spacetimes under consideration, the dimensionless impact parameter $\bar{b}$ is given by (see Eq. \ref{eq:impact_turning})
\begin{equation}
\bar{b}=\frac{1}{u_{tp}\sqrt{f_i(u_{tp})}}.
\label{eq:impact_junction1}
\end{equation}
Note that when $r_{tp}=R_b$, $u_{tp}=1$. Therefore, if $u_{tp}<1$, then the photon does not enter the interior of the JMN metric. In that case, the deflection is given by Eqs. \eqref{eq:deflection_junction1} and \eqref{eq:impact_junction1} with $f_i(r)$ and $g_i(r)$ given by
the exterior Schwarzschild metric. However, if the photon does enter the JMN metric and has its turning point in the interior ($r_{tp}<R_b$, i.e., $u_{tp}>1$), then the deflection angle can be written as \citep{Sahu}
\begin{eqnarray}
\alpha &=& 2\int_0^{1}\frac{1}{\sqrt{g_0}}\frac{1}{\sqrt{\frac{1}{\bar{b}^2f_0(u)}-u^2}}du \nonumber \\
&& + 2\int_1^{u_{tp}}\frac{1}{\sqrt{g_{1,2}}}\frac{1}{\sqrt{\frac{1}{\bar{b}^2f_{1,2}(u)}-u^2}}du-\pi,
\label{eq:deflection_junction2}
\end{eqnarray}
where $f_{1,2}$ refer to the JMN-1 or JMN-2 models, respectively, and the impact parameter $\bar{b}$ is given by Eq. \eqref{eq:impact_junction1}.  The first term in Eq. \eqref{eq:deflection_junction2} is the contribution from the exterior Schwarzschild geometry and the second term is that from the interior JMN metric, $\alpha_{\mbox{JMN1}}$ or $\alpha_{\mbox{JMN2}}$.

Because of its simple form, here we focus on the JMN-1 naked singularity. As discussed in Sec \ref{sec:shadow_boundary}, a photon sphere exists for $M_0\geq 2/3$ and the photon sphere lies in the exterior Schwarzschild geometry. As a result, all the photons which participate in the image formation have their turning points outside of the photon sphere. Therefore, in this case, there is no difference in lensing behavior between the JMN-1 naked singularity and the Schwarzschild black hole. On the other hand, since there is no photon sphere for $M_0<2/3$, photons may enter the interior of the JMN-1 spacetime and experience a turning point because of the infinite potential barrier at the singularity. Therefore, for this range of $M_0$, there is a clear distinction between the lensing behavior of the JMN-1 naked singularity and that of the Schwarzchild black hole.

For $M_0<2/3$, the contribution of the JMN-1 spacetime to the deflection angle, $\alpha_{\mbox{JMN1}}$, can be obtained analytically by a change of variables to $z=u^{(2-3M_0)/2(1-M_0)}$. We then obtain 
\begin{eqnarray}
\alpha_{\mbox{JMN1}} &=& 2\int_1^{u_{tp}}\frac{du}{\sqrt{g_{1}}}\frac{1}{\sqrt{\frac{1}{ \bar{b}^{2} f_{1}(u)}-u^2}} \nonumber \\
&=& \frac{4\sqrt{1-M_0}}{2-3M_0} \int_1^{z_{tp}} \frac{dz}{\sqrt{\frac{1}{\bar{b}^{2}(1-M_0)}-z^2}} \nonumber\\
&=& \frac{2\sqrt{1-M_0}}{2-3M_0}\left[\pi-2\sin^{-1}\left(\frac{r_{tp}}{R_b}\right)^{\frac{2-3M_0}{2(1-M_0)}}\right],
\end{eqnarray}
where $r_{tp}\leq R_b$. 

\begin{figure}
\centering
\includegraphics[scale=0.67]{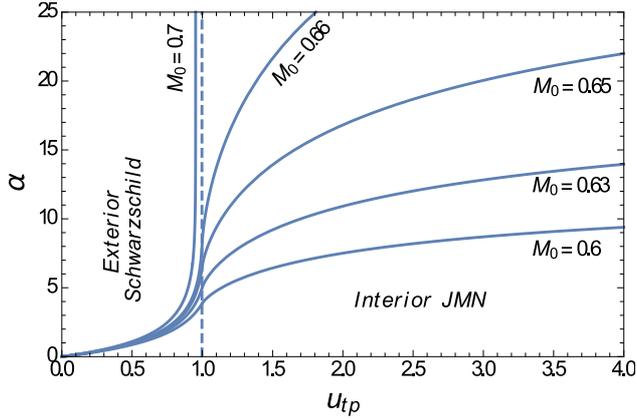}
\caption{
Deflection angle $\alpha$ as a function of $u_{tp}$ of light rays for the JMN-1 naked singularity matched with an exterior Schwarzschild geometry. The vertical dashed line shows the boundary between the two geometries.}
\label{fig:bending_JMN1}
\end{figure}

The analytical expression of the contribution due to the Schwarzschild geometry in the exterior of the JMN-1 model is the same as that of a Schwarzschild black hole and can be found in \citet{sch_lensing1}. Figure (\ref{fig:bending_JMN1}) shows a plot of the deflection angle as a function of $u_{tp}$. Since a photon sphere exists for $M_0\geq 2/3$, the deflection angle diverges as the turning point approaches the photon sphere. This divergence is logarithmic \citep{bozza}. Therefore, theoretically, there will be an infinite number of images just outside the photon sphere. 

For the JMN-1 naked singularity model with $M_0<2/3$, although there is no photon sphere, the deflection angle can still be large because, depending on the impact parameter, light rays may wind around the singularity several times. Due to this large bending, there can be many relativistic rings even for $M_0<2/3$.  

In the following, for simplicity, we assume that the observer, the lens, and the distant point light source are all aligned. We also consider that the observer and the light source are far away from the
lens. Therefore, in the observer's sky, the relativistic images will be concentric rings (known as relativistic Einstein rings) of radii given by the corresponding impact parameters $b(r_{tp})$.  These
impact parameter values $b(r_{tp})$ can be obtained by solving $\alpha\simeq 2\pi n$, where $n$ is the ring number \citep{bozza}.

\begin{figure*}
\centering
\subfigure[$M_0=0.6$, JMN-1 naked singularity]{\includegraphics*[scale=0.5]{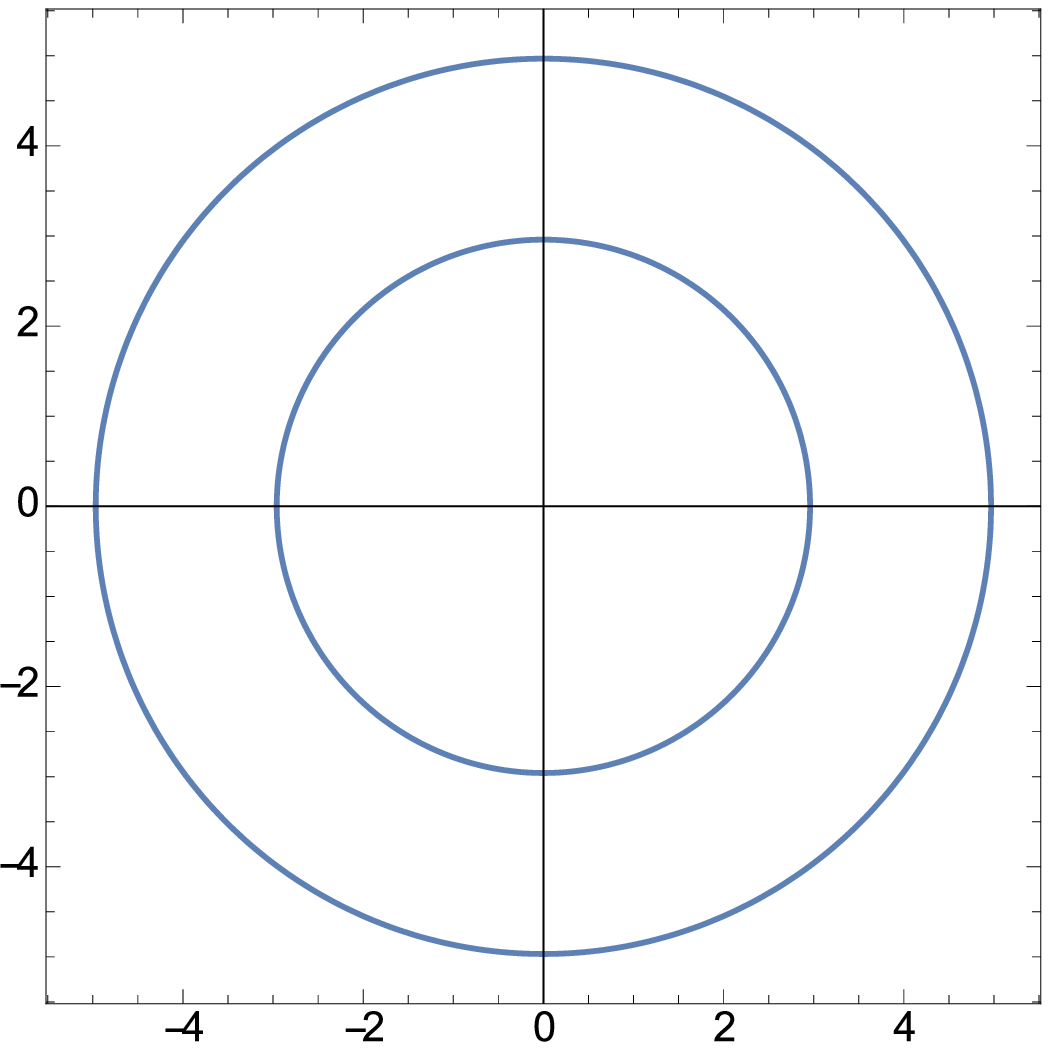}}\hspace{0.1cm}
\subfigure[$M_0=0.63$, JMN-1 naked singularity]{\includegraphics*[scale=0.5]{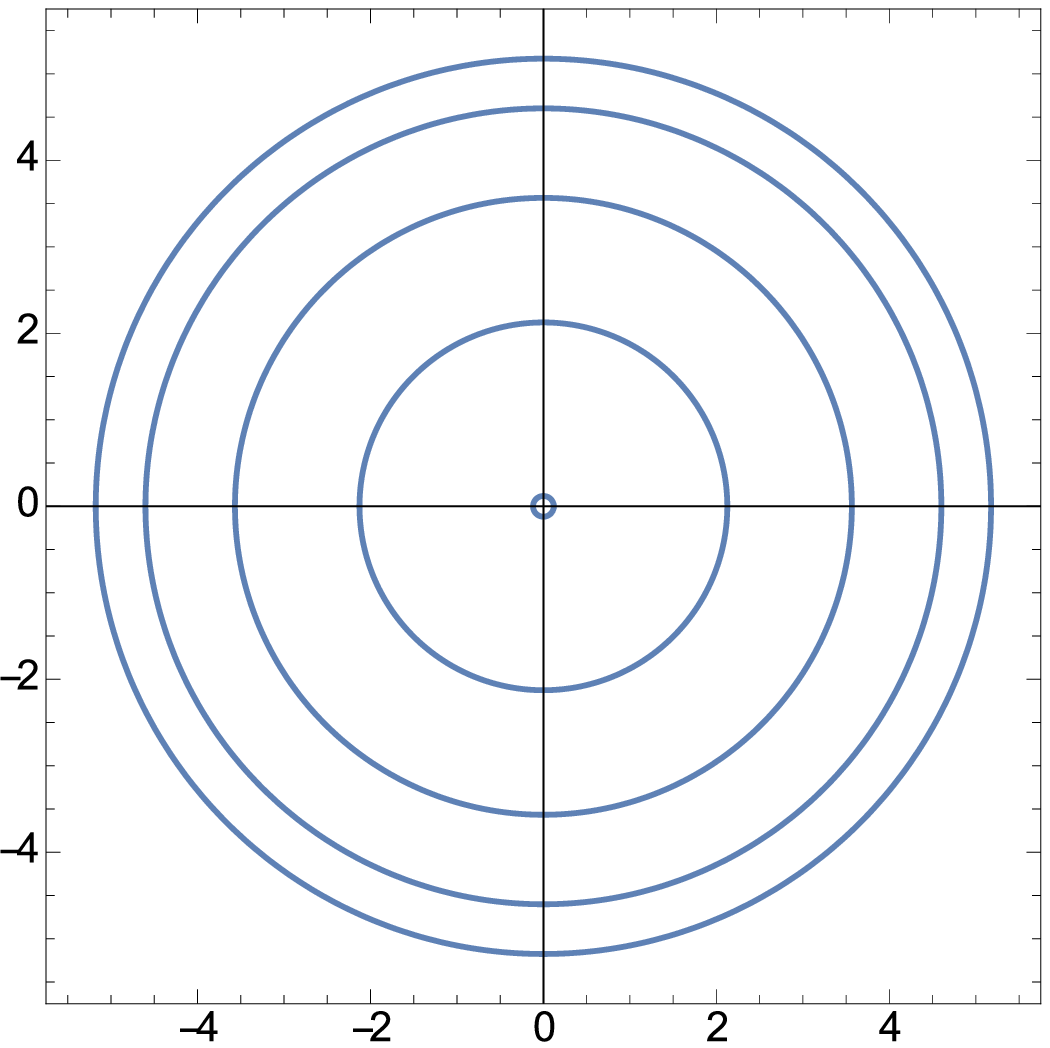}}
\subfigure[$M_0=0.65$, JMN-1 naked singularity]{\includegraphics*[scale=0.5]{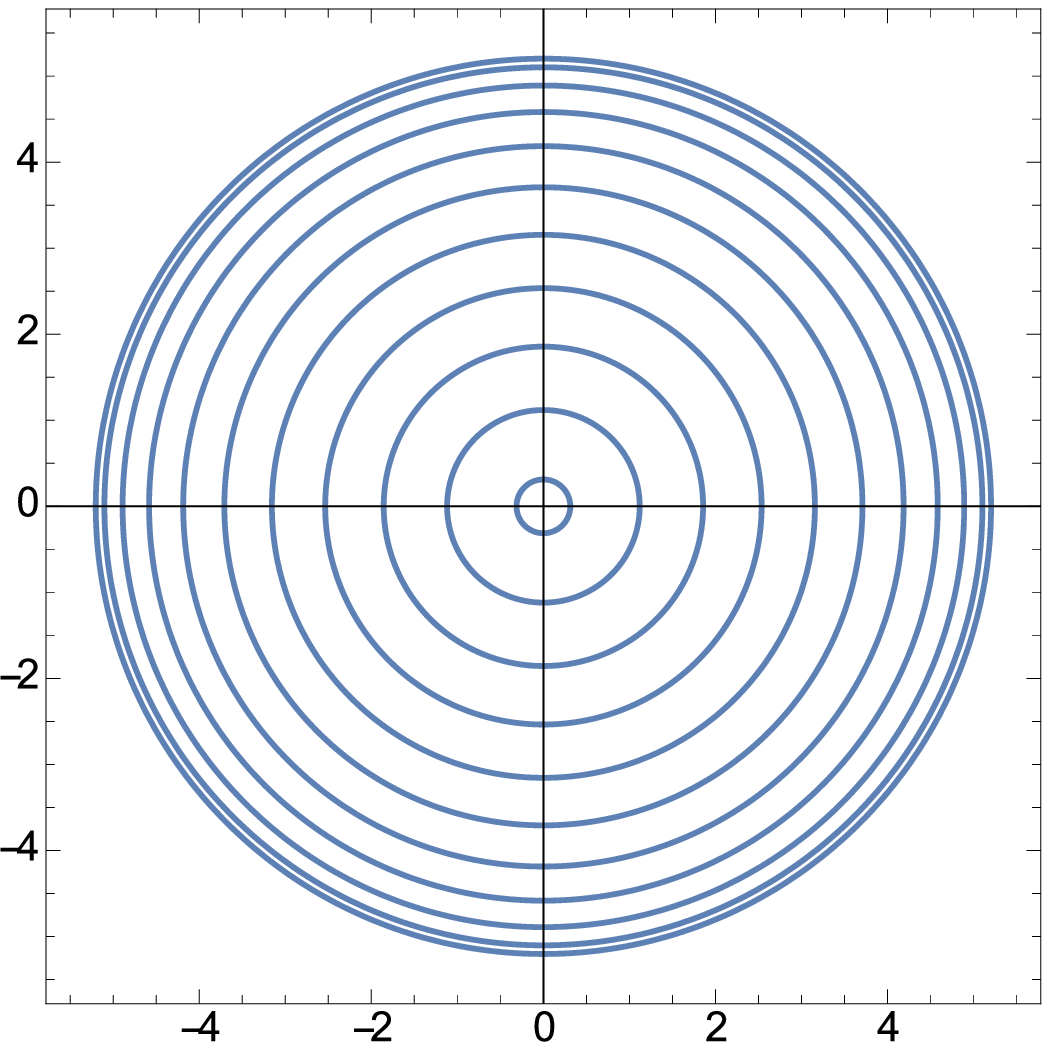}}\hspace{0.1cm}
\subfigure[$M_0=0.66$, JMN-1 naked singularity]{\includegraphics*[scale=0.5]{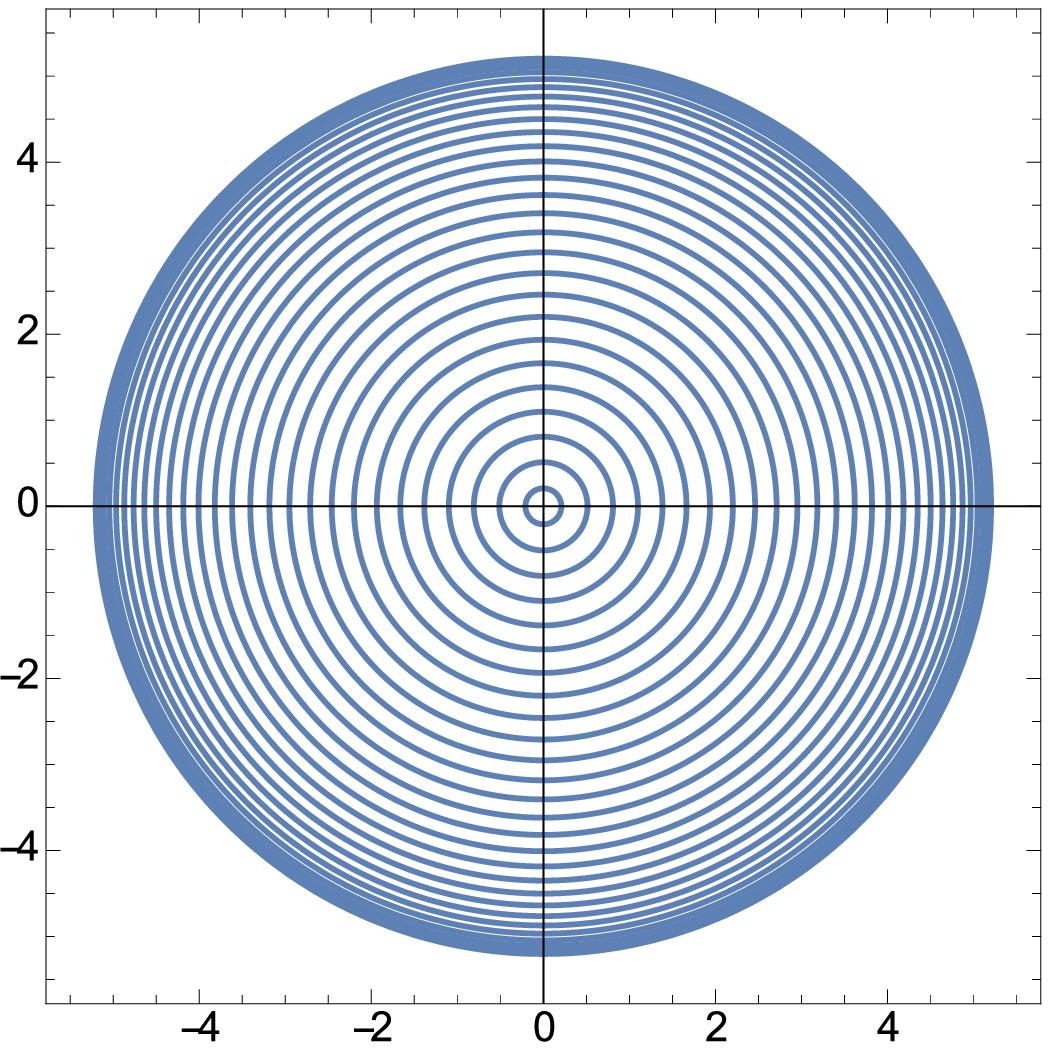}}
\subfigure[$M_0=0.7$, JMN-1 naked singularity]{\includegraphics*[scale=0.5]{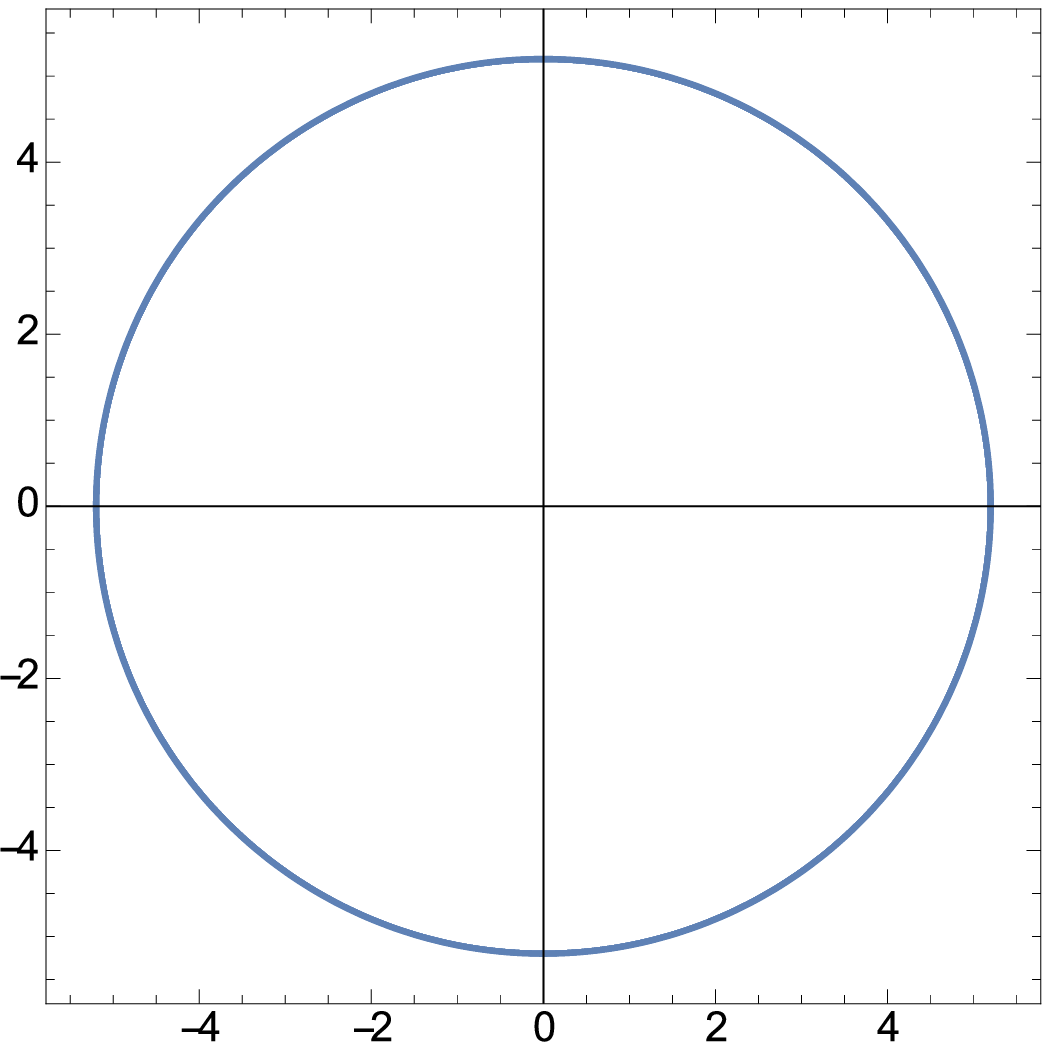}}\hspace{0.1cm}
\subfigure[Schwarzschild black hole]{\includegraphics*[scale=0.5]{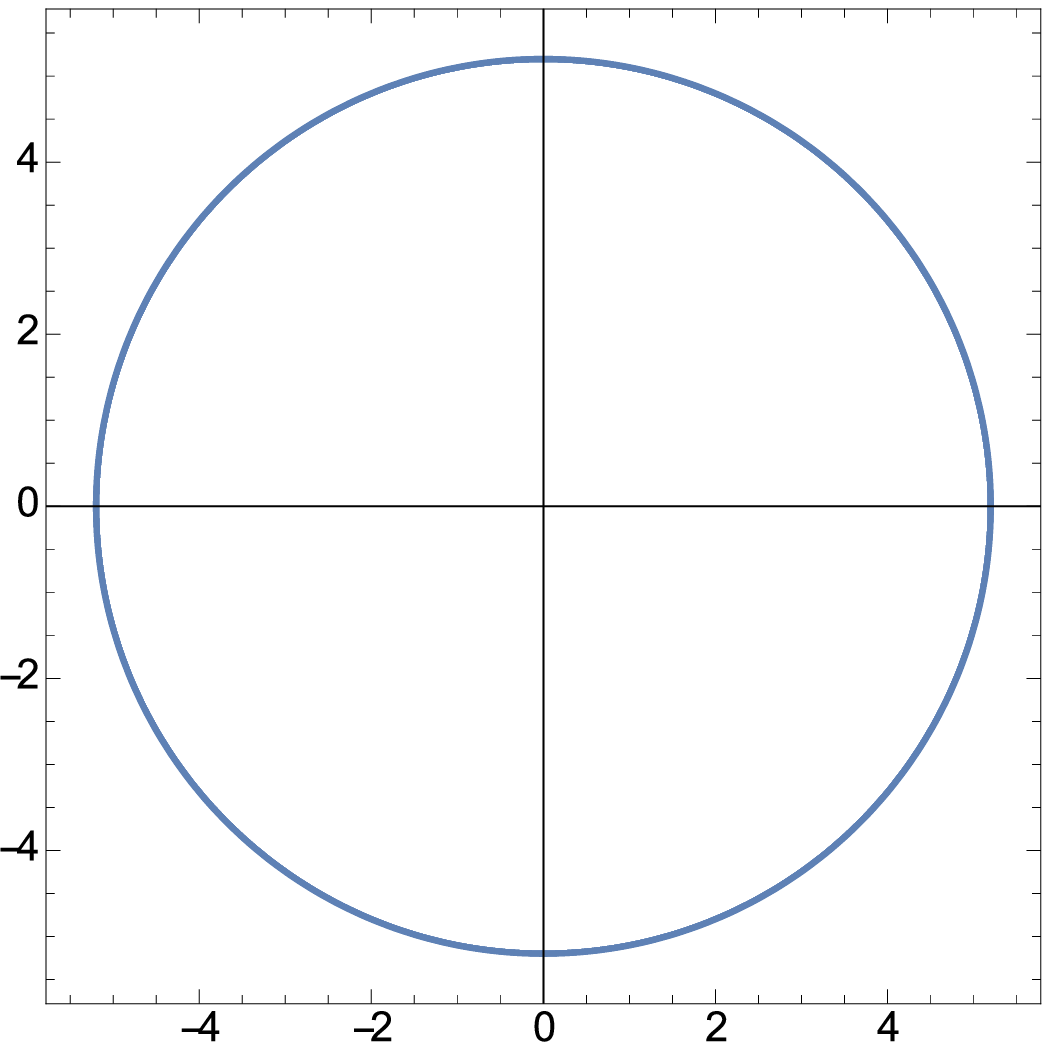}}
\caption{
Relativistic Einstein ring images due to gravitational lensing by (a-e) the JMN-1 naked singularity with different values of $M_0$, and (f) the Schwarzschild black hole. The axes are in units of $M$.}
\label{fig:rings_JMN1}
\end{figure*}

Figure \ref{fig:rings_JMN1} shows the relativistic Einstein rings in the observer's sky. In the case of the JMN-1 naked singularity with $M_0\geq 2/3$ and the Schwarzschild black hole, all the relativistic
images are clumped together outside the photon sphere, which forms the outer boundary of the shadow. The radius of the innermost image in the observer's sky is given by the minimum critical impact parameter $b_{ph}$. Photons with impact parameter less than $b_{ph}$ are absorbed by the photon sphere. Hence, in this case, there is a shadow and many relativistic images clumped together just outside the edge of the shadow. 

As an aside we note that, besides the relativistic Einstein rings discussed here, there is a standard Einstein ring formed as a result of weak deflection of light (weak deflection occurs when $M/r_{tp}\ll
1$). For all physically reasonable $M_0$ values, weak deflection and the traditional Einstein ring occur in the exterior Schwarzschild geometry. Therefore, the traditional Einstein ring of the JMN-1 naked singularity will be the same as that due to the Schwarzschild black hole (differences may arise when $M_0$ is uninterestingly small).

In the case of the JMN-1 naked singularity with $M_0<2/3$, there is no photon sphere and hence there is no capture of photons. As a result, we have many distinct rings corresponding to different relativistic
images. The density of relativistic images increases as $M_0$ approaches $2/3$. This is illustrated in Figure \ref{fig:rings_JMN1}. 

The discussion so far is for a single point source aligned perfectly behind the lens. However, in realistic situations, we may have many light sources in different directions and at different distances
around the lens. The angular positions of the relativistic images formed due to each source will be different. Therefore, in the observer's sky, there will be numerous relativistic images, which
might fill the gaps between the relativistic images shown in Fig. \ref{fig:rings_JMN1}. Hence, we may have a smooth continuous image.  A similar situation occurs when the black hole or the JMN naked singularity is surrounded by an optically thin emission region, as we discuss in the next section.

\section{Shadows and images of optically thin emission regions surrounding black holes and naked singularities}
\label{sec:shadow}
The previous two sections dealt with distant sources of radiation, far behind the lensing compact object.  Here we consider an optically thin, radiating, accretion flow surrounding the compact object and compute the observed image. The difference is that radiation is now emitted over an extended volume near the compact object, including regions inside the photon sphere.

The observed specific intensity $I_{\nu_0}$ (usually measured in erg~s$^{-1}$~cm$^{-2}$~str$^{-1}$~Hz$^{-1}$) at the observed photon frequency $\nu_o$ at the point $(X,Y)$ in the observer's sky is given by \citep{intensity_formula1,intensity_formula2}, 
\begin{equation}
I_{\nu_{o}}(X,Y) = \int_\gamma g^3\; j (\nu_{\rm e}) dl_{\rm prop},
\label{eq-I}
\end{equation}
where $\nu_e$ is the emitted frequency, $g = \nu_{o}/\nu_{e}$ is the redshift factor, $j(\nu_e)$ is the emitter's rest-frame emissivity per unit volume, $dl_{prop}=-k_\alpha u^\alpha_e d\lambda$ is the infinitesimal proper length in the rest frame of the emitter, $k^\mu$ is the four-velocity of the photons, $u^\mu_{e}$ is the four-velocity of the emitter, and $\lambda$ is the affine parameter along the photon path $\gamma$. The subscript $\gamma$ on the integral means that the integration is evaluated along an observed photon path $\gamma$. The redshift factor $g = \nu_{o}/\nu_{e}$ is given by,
\begin{equation}
g = \frac{k_\alpha u^\alpha_{o}}{k_\beta u^\beta_{e}},
\end{equation}
where $u^\mu_{o} = (1,0,0,0)$ is the four-velocity of the distant observer (who is at infinity).

In the spirit of the simple spherically-symmetric spacetimes we are investigating, we consider a correspondingly simple model for the accreting gas. We assume that the gas is in radial free fall \citep{intensity_formula2}, with its four-velocity given by,
\begin{equation}
u^t_{e} = \frac{1}{f_i(r)}, \;\; 
u^r_{e} = - \sqrt{\frac{g_i(r)}{f_i(r)}\left[1 - f_i(r)\right]}, \;\;
u^\theta_{\rm e} = u^\phi_{\rm e} = 0.
\label{eq:free_fall_velocity}
\end{equation}
The four-velocity $k^\mu$ ($=\dot{x}^\mu$) of the photons was already obtained previously. In the subsequent calculations, we will need the following expression,
\begin{equation}
\frac{k^r}{k^t} =\pm f_i(r) \sqrt{g_i(r)\left[\frac{1}{f_i(r)} -\frac{b^2}{r^2}\right]},
\end{equation}
where the sign $+$($-$) is when the photon moves away from (approaches towards) the massive object. The redshift function $g$ is thus given by,
\begin{equation}
g=\frac{1}{\frac{1}{f_i(r)}-\frac{k_r}{k_t}\sqrt{\frac{g_i(r)}{f_i(r)}(1-f_i(r))}}.
\end{equation}
For the specific emissivity, we assume the following simple model \citep{intensity_formula2} in which the emission is monochromatic with emitter's rest-frame frequency $\nu_\star$, and the emission has a $1/r^2$ radial profile:
\begin{equation}
j(\nu_{\rm e}) \propto \frac{\delta(\nu_e - \nu_\star)}{r^2},
\end{equation}
where $\delta$ is the Dirac delta function.  Finally, the proper length in the emitter frame is given by
\begin{equation}
dl_{\rm prop}=-k_\alpha u^\alpha_e d\lambda = -\frac{k_t}{g k^r } dr.
\end{equation}
Integrating Eq. \eqref{eq-I} over all the observed frequencies, we obtain the observed photon intensity \citep{intensity_formula2}
\begin{equation}
I_{obs} (X,Y) \propto -\int_\gamma \frac{g^3 k_t dr}{r^2 k^r}.
\end{equation}

\begin{figure*}
\centering
\subfigure[~$M = 1.0$, Schwarzschild black hole]{\includegraphics[scale=0.60]{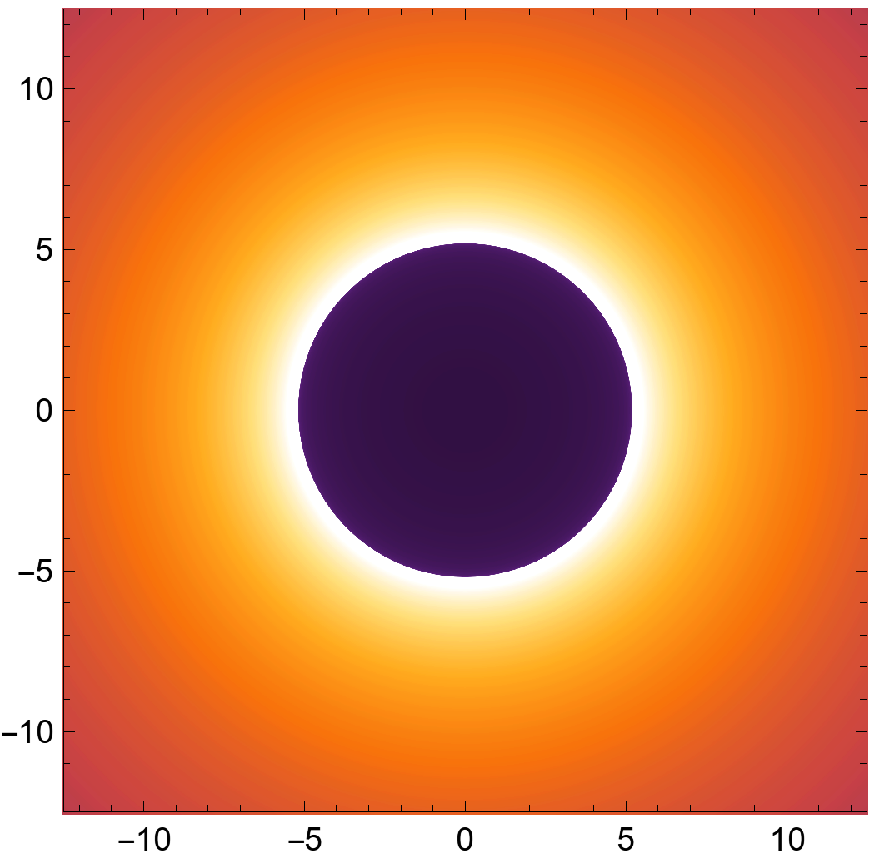}}\hspace{0.1cm}
\subfigure[~$M_0 = 0.7$, JMN-1 naked singularity]{\includegraphics[scale=0.60]{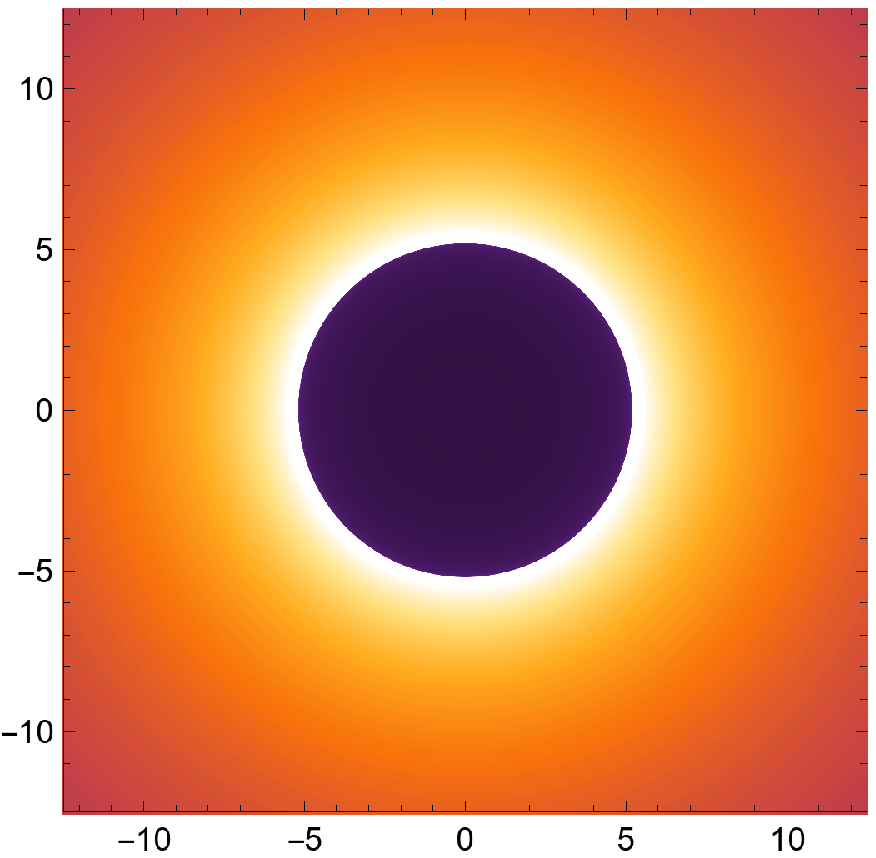}}\hspace{0.1cm}
\subfigure[~$M_0 = 0.6$, JMN-1 naked singularity]{\includegraphics[scale=0.60]{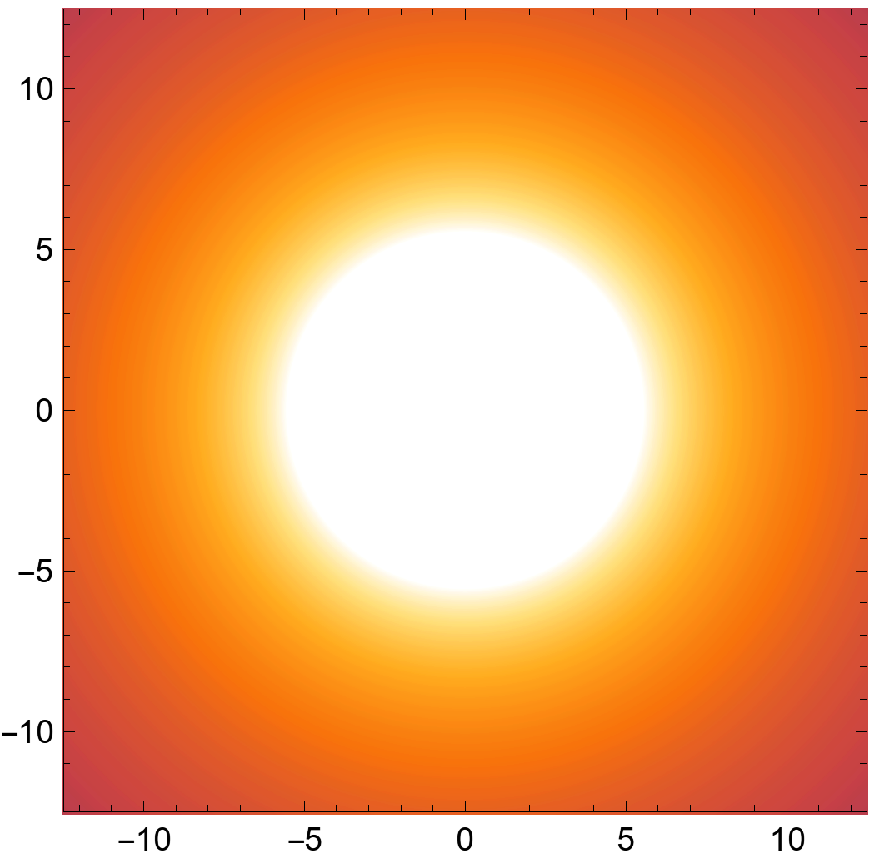}}
\subfigure[~$M = 1.0$, Schwarzschild black hole]{\includegraphics[scale=0.53]{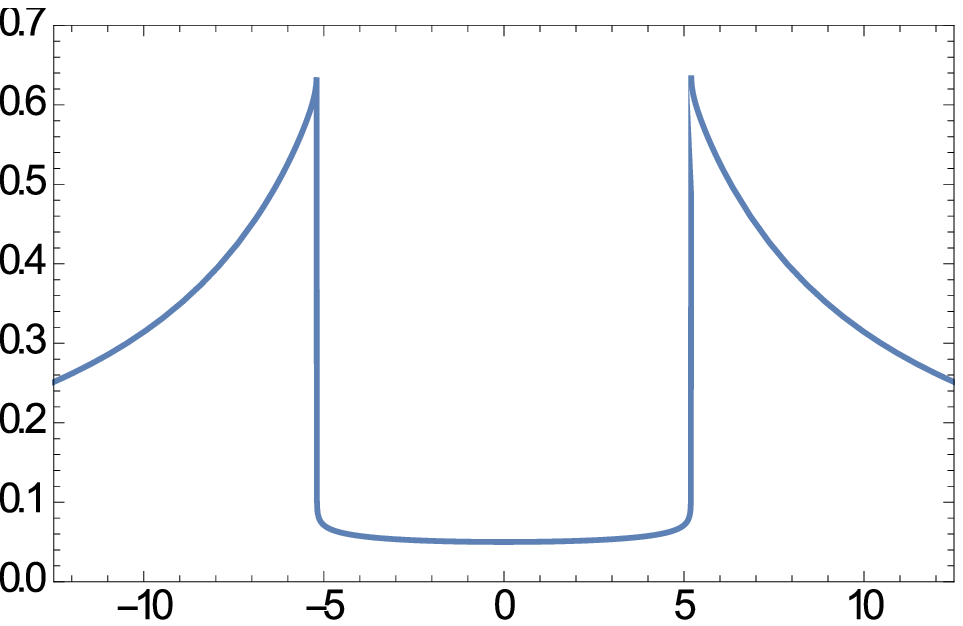}}\hspace{0.1cm}
\subfigure[~$M_0 = 0.7$, JMN-1 naked singularity]{\includegraphics[scale=0.53]{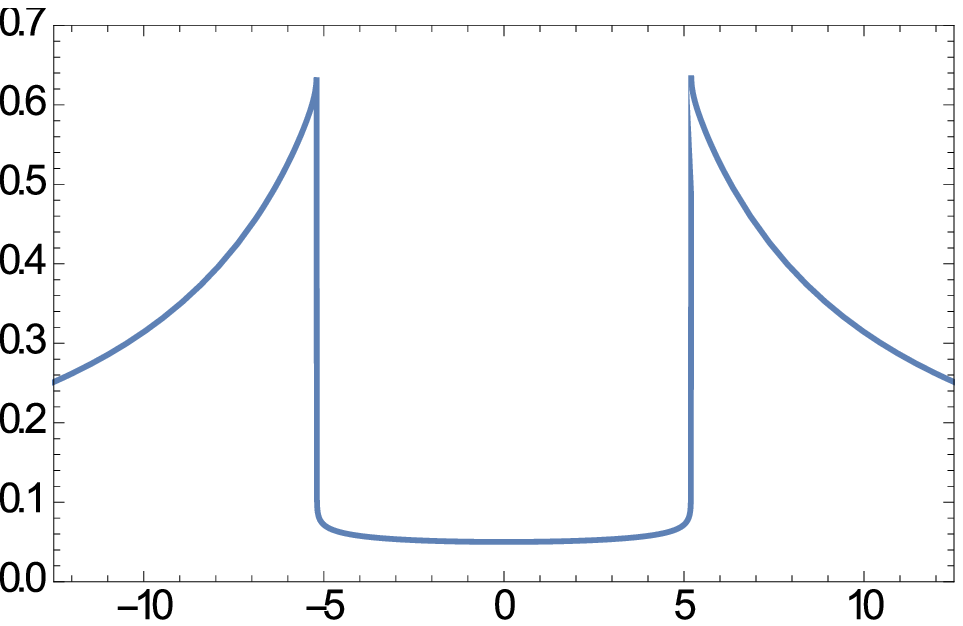}}\hspace{0.1cm}
\subfigure[~$M_0 = 0.6$, JMN-1 naked singularity]{\includegraphics[scale=0.53]{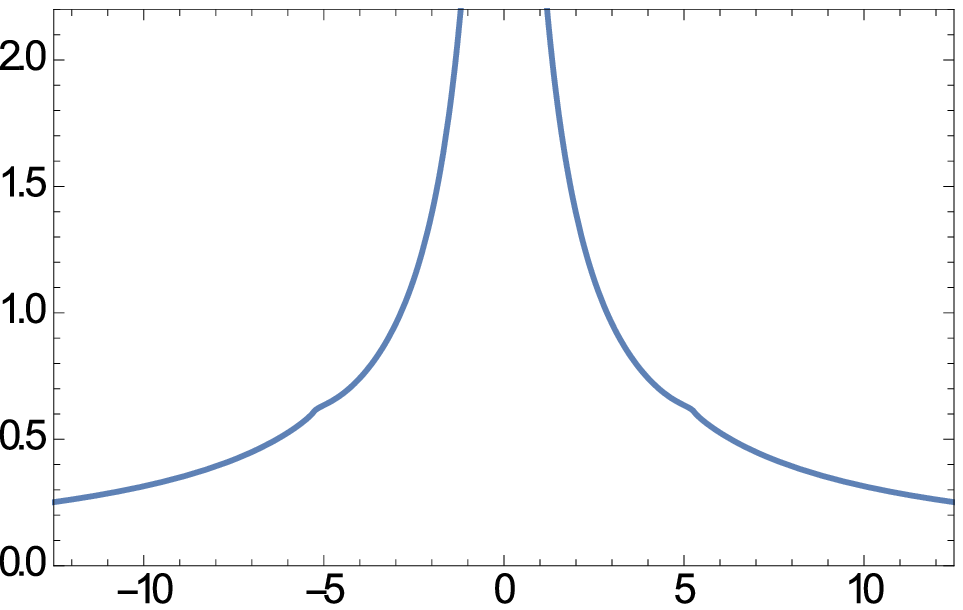}}
\caption{
The upper row shows the image of an optically thin emission region surrounding the Schwarzschild black hole (a) and the JMN-1 naked singularity for $M_0=0.7$ (b) and $M_0=0.6$ (c). The corresponding intensity distributions as a function of the impact parameter are shown in the lower row. All spatial coordinates are in units of $M$.}
\label{fig:JMN1}
\end{figure*}

\begin{figure*}
\centering
\subfigure[~$M = 1.0$, Schwarzschild black hole]{\includegraphics[scale=0.6]{image_BH_new.eps}}\hspace{0.1cm}
\subfigure[~$\lambda = 0.8$, JMN-2 naked singularity]{\includegraphics[scale=0.6]{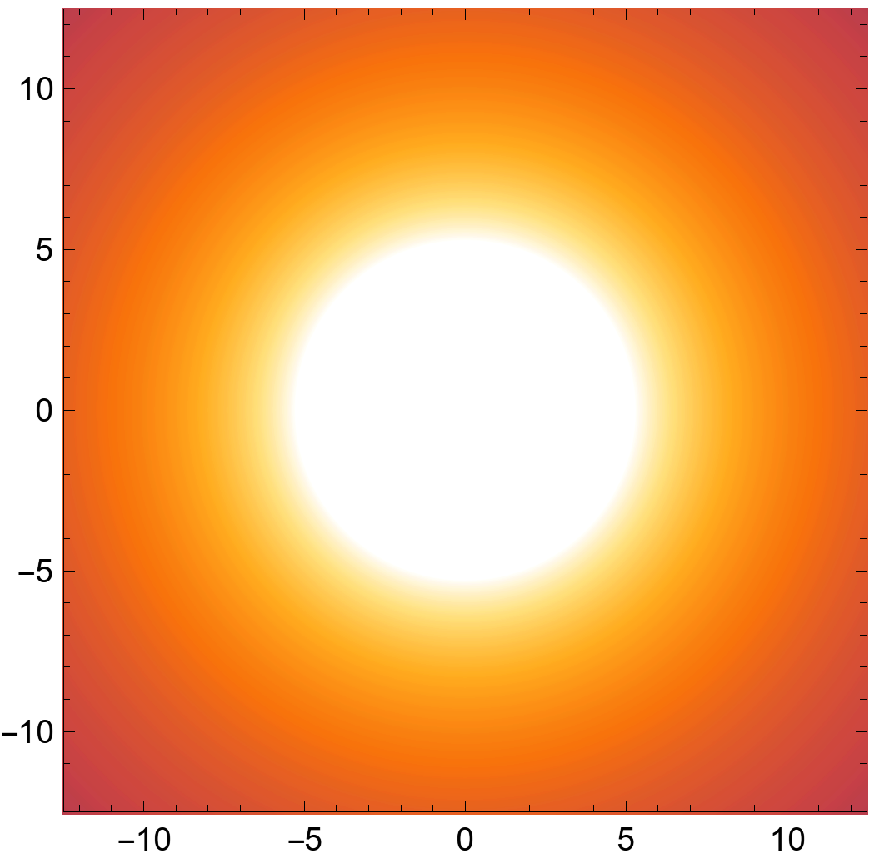}}\hspace{0.1cm}
\subfigure[~$\lambda = 0.4$, JMN-2 naked singularity]{\includegraphics[scale=0.6]{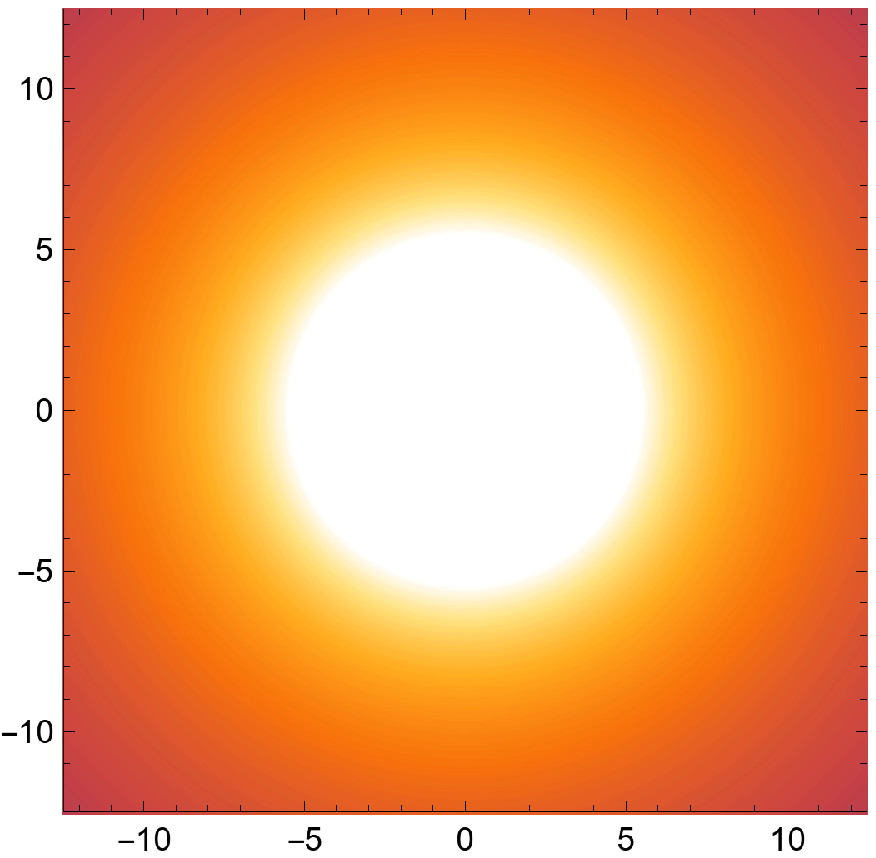}}
\subfigure[~$M = 1.0$, Schwarzschild black hole]{\includegraphics[scale=0.53]{intensity_BH.eps}}\hspace{0.1cm}
\subfigure[~$\lambda = 0.8$, JMN-2 naked singularity]{\includegraphics[scale=0.53]{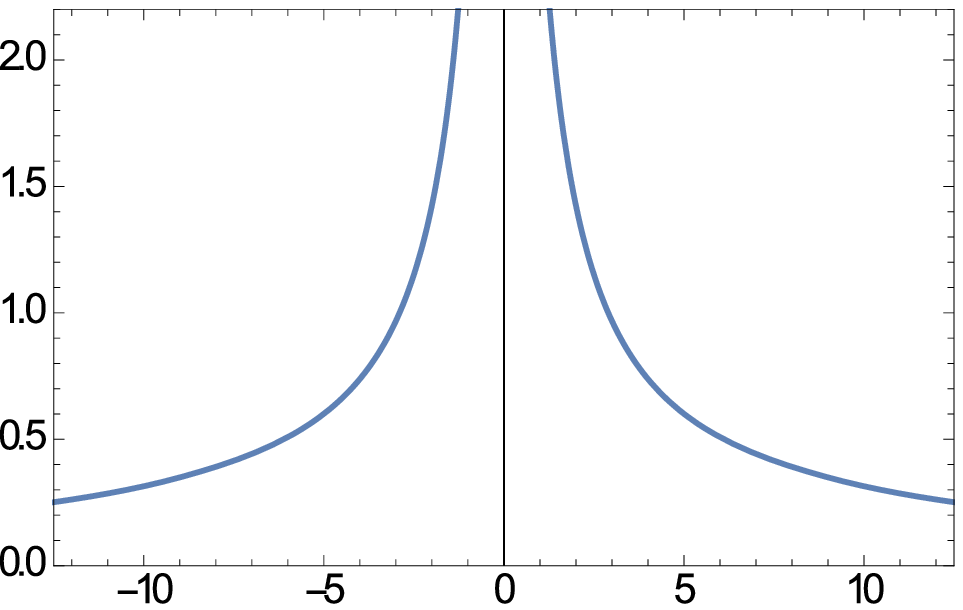}}\hspace{0.1cm}
\subfigure[~$\lambda = 0.4$, JMN-2 naked singularity]{\includegraphics[scale=0.53]{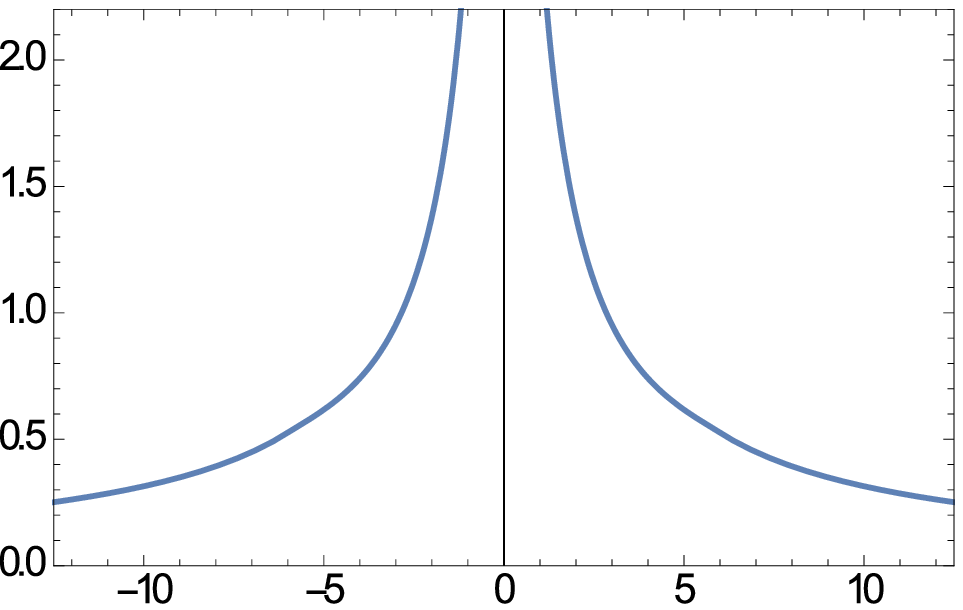}}
\caption{
The upper row shows the image of an optically thin emission region surrounding the Schwarzschild black hole (a) and the JMN-2 naked singularity for $\lambda=0.8$ (b) and $\lambda=0.4$ (c). The corresponding intensity distributions as a function of the impact parameter are shown in the lower row.  All spatial coordinates are in units of $M$.}
\label{fig:JMN2}
\end{figure*}

Note that the intensity map in the observer's sky will be circularly symmetric, with the impact parameter $b$ of any equi-intensity circle given by $X^2+Y^2=b^2$.  Figures \ref{fig:JMN1} and \ref{fig:JMN2} show intensity maps of the image of the above model accretion flow for the Schwarzschild black hole and the two JMN naked singularities.

We now note the qualitative differences in the shadows and images produced by the different models. As expected, the Schwarzschild black hole always casts a shadow (Fig. \ref{fig:JMN1}a), though we should point out that the intensity inside the shadow does not quite go to zero as in the previous sections but has a small finite value (Fig. \ref{fig:JMN1}d). This difference is because the accretion flow emits radiation inside the photon sphere and a small fraction of this radiation is able to escape to infinity.

In the case of the JMN-1 naked singularity, if the model has a photon sphere ($M_0\geq2/3$, $R_b \leq 3M$), then its shadow and image (Fig. \ref{fig:JMN1}b) mimic those of the Schwarzschild black hole. However, if the JMN-1 naked singularity does not have a photon sphere, then it casts a ``full-moon" image (Fig. \ref{fig:JMN1}c), which is remarkably different from the images in Figs. \ref{fig:JMN1}a and \ref{fig:JMN1}b.  
Such a difference, if observationally detected, could greatly help distinguish a naked singularity from a black hole.

The JMN-2 naked singularity model does not have a photon sphere for any allowed value of the parameter $\lambda$. Therefore, the image in this case is always a full-moon, as illustrated in Figs. \ref{fig:JMN2}b and \ref{fig:JMN2}c.

Our results suggest that, though a naked singularity that has a photon sphere cannot be distinguished from a black hole through observations of the shadow or image, a naked singularity that does not have any photon sphere can be. Note that naked singularities without photon spheres arise when physically realistic collapse models are considered, such as the JMN-2 model \citep{JMN_pressure}.

Photons emitted from the close vicinity of either JMN naked singularity are highly redshifted. However, this is compensated by the fact that the rate of emission of photons from the accretion flow in the vicinity of the singularity is large. The redshift of photons emitted in the forward direction of the accretion flow is smaller than those emitted in the backward direction. Therefore, although the contribution of ``backward photons'' may be highly suppressed, ``forward photons'' turn and escape (in the absence of a photon sphere), contributing to the intensity profile. The center of the image in the observer's sky corresponds to photons with zero impact parameter. Such photons originate from the close vicinity of the singularity as well as from the accreting matter in between the singularity and the observer, along the line joining them. We note also that the redshift is irrelevant when considering the formation of shadows and full-moon images due to distant extended sources of light, as discussed in \S\ref{sec:shadow_boundary} and \S\ref{sec:lensing}, since the blueshift suffered by a photon in going from the source to the turning point nearly cancels out the redshift suffered by it in going from the turning point to a faraway observer. In that case, the total redshift suffered by the photon is simply determined by the relative redshift between the source and observer's positions. When both the light source and the observer are at large distances, this redshift is negligible.

\section{Shadows and images of more realistic accretion flows around black holes and naked singularities}
\label{sec:shadow2}

We now describe a more realistic model of the accretion flow with
several improvements: (i) we consider a physically motivated
emissivity prescription, (ii) we analyze the spectrum of the
radiation, and (iii) we avoid the assumption of optically thin
emission.  As we show, the results are similar to those obtained in
\S\ref{sec:shadow}.

\subsection{The model}

With a view to specializing to the case of the Galactic Center compact
object Sagittarius A$^*$ (Sgr A$^*$), we consider a specific value for
the mass of the central object: $M=4\times10^6M_\odot$.  As in
\S\ref{sec:shadow}, we assume a spherically symmetric accretion flow,
except that we set up the dynamics as in the Bondi accretion model
\citep{bondi}.  Thus, we assume that the compact object is embedded in
a uniform external medium with a temperature $T_\infty$ and density
$\rho_\infty$. We choose $T_\infty=10^7$\,K, as appropriate for Sgr
A$^*$. For this choice, the Bondi radius, i.e., the transition radius
where the flow changes in character from a uniform external medium to
a freely-falling inner accretion flow, is $r_B\approx 10^6M$. We keep
$\rho_\infty$ as a free parameter which we adjust (thereby tuning the
mass accretion rate) such that the luminosity of the resulting
accretion flow in the sub-millimeter band matches the observed flux of
Sgr A$^*$.  Finally, in the spirit of the Bondi model, and in keeping
with \S\ref{sec:shadow}, we take the velocity profile of the accreting
gas to be given by Eq.~\eqref{eq:free_fall_velocity}, using the
appropriate $f_i(r)$ and $g_i(r)$ for each model. However, we modify
the radial velocity profile at large radii so that the velocity
transitions from the standard free-fall scaling, $v_r \propto
r^{-1/2}$, at radii inside the Bondi radius to $v_r \propto r^{-2}$
outside the Bondi radius (as required for a constant mass accretion
rate with a uniform gas density at large radii).

We assume that the accreting gas radiates thermal synchrotron and
bremsstrahlung, and that the emitted radiation is Compton-scattered as
it propagates out of the system. The radiation is treated via a
complete radiative transfer model using the transfer code HEROIC
\citep{heroic1,heroic2}, with the relativistic enhancements described
in \citet{heroic_ulx}. In this code, a large number of ray directions
is considered at each point in the accretion flow and the relativistic
radiative transfer equation, which considers both emission and
absorption, is solved for each ray over a grid of frequencies
extending from $\nu=10^8$\,Hz to $10^{24}$\,Hz. HEROIC was originally
written for the Kerr spacetime, and all previous applications were
restricted to that spacetime.  For the present application, the code
was generalized to handle the JMN-1 and JMN-2 spacetimes as well.

The radiative transfer computations enable us to compute the
luminosity and radiative spectrum of the emerging radiation for each
model.  In addition, they also provide the net cooling (if emission
dominates) or heating (if absorption dominates) of the accreting
gas. We include this cooling/heating information in the energy
equation of the accreting gas to solve for the temperature profile
$T(r)$ of the flow\footnote{For simplicity, we assume that the gas is
  a single-temperature plasma, although it is likely that the
  accreting gas in Sgr A$^*$ is a two-temperature plasma
  \citep{yuan_narayan14}}. In other words, the only temperature
information we input to the model is the boundary condition at
infinity ($T_\infty=10^7$\,K), which sets the location of the Bondi
radius.  The temperature everywhere else is obtained self-consistently
as part of the solution.

The numerical computations are done on a uniform grid in $\log r$,
with 20 points per decade. The grid extends from an outer radius
$r_{\rm max} = 10^{6.5}M$ (a factor of a few larger than the Bondi
radius) down to an inner radius $r_{\rm min}$. In the case of the
Schwarzschild black hole, we choose $r_{\rm min}$ to be just outside
the horizon, specifically, $\log r_{\rm min} = 0.35$.  We assume
absorbing boundary conditions at the inner edge of the grid, i.e., any
radiation that crosses the horizon is lost from the system.  For the
two naked singularity models, we would ideally like to set $r_{\rm
  min}=0$, but this is not possible 
because of our use of a logarithmic grid. Hence, we use a small non-zero value,
$r_{\rm min} = 10^{-4}M$, again assuming absorbing boundary
conditions.

\subsection{Spectra and Temperature Profiles}

\begin{figure}
\centering
\includegraphics[scale=1.0]{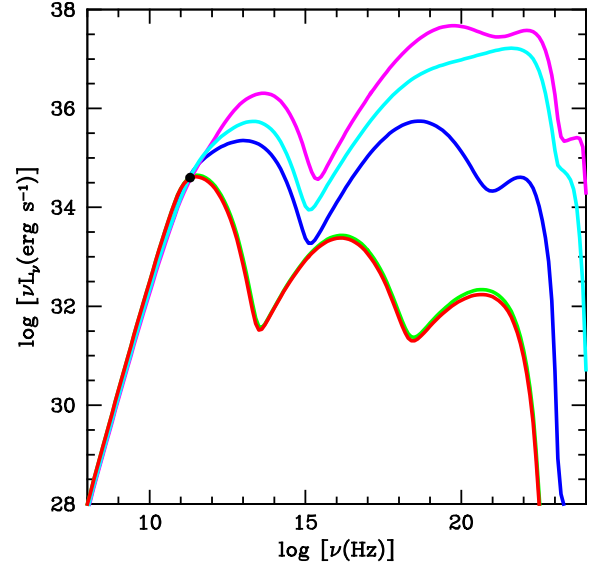}
\vspace{-2.5cm}
\caption{Spectra of models with a Schwarzschild black hole (red),
  JMN-1 naked singularity with $M_0 = 0.7$ (green, under red) and
  $M_0=0.6$ (blue), and JMN-2 naked singularity with $\lambda=0.8$
  (magenta) and $\lambda=0.4$ (cyan).}
\label{fig:spectra}
\end{figure}

\begin{figure}
\centering
\includegraphics[scale=1.0]{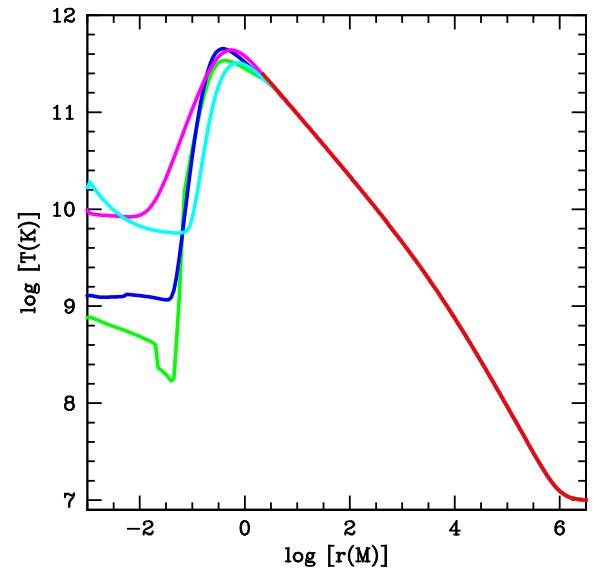}
\vspace{-2.5cm}
\caption{Radial temperature profiles of models with a Schwarzschild
  black hole (red), JMN-1 naked singularity with $M_0 = 0.7$ (green)
  and $M_0=0.6$ (blue), and JMN-2 naked singularity with $\lambda=0.8$
  (magenta) and $\lambda=0.4$ (cyan).}
\label{fig:temperatures}
\end{figure}

Figure \ref{fig:spectra} shows spectra corresponding to five different models: Schwarzschild black hole (red curve), JMN-1 naked singularity with $M_0 = 0.7$ (green) and $M_0=0.6$ (blue), and JMN-2 naked singularity with $\lambda=0.8$ (magenta) and $\lambda=0.4$ (cyan). In each model, the mass accretion rate has been adjusted (by varying the density $\rho_\infty$ of the external medium) so as to give the same luminosity, $\nu L_\nu = 10^{34.6}~{\rm erg\,s^{-1}}$ at $\nu=200$\,GHz (indicated by the black dot), as seen by an observer at infinity. This is approximately the luminosity of Sgr A$^*$.

As explained in previous sections, of the five spacetime models under consideration, only two have photon spheres, viz., the Schwarzschild black hole and the JMN-1 spacetime with $M_0=0.7$. Not surprisingly, these two models have nearly identical spectra. The primary peak at $10^{11}$\,Hz is due to thermal synchrotron radiation from hot electrons at radii near the photon sphere. The other peaks are the result of Compton scattering, with a small contribution from bremsstrahlung in the last peak. 

The Schwarzschild model and the JMN-1 model with $M_0=0.7$ are much less luminous than the other three models (JMN-1 $M_0=0.6$, JMN-2 $\lambda=0.8$, JMN-2 $\lambda=0.4$). The latter three spaceimes lack photon spheres and therefore allow radiation to escape more easily from the interior. As a result, they appear to be substantially more luminous, by orders of magnitude, for an observer at infinity.

Figure \ref{fig:temperatures} shows the temperature as a function of radius for the same five models. All have essentially the same profile at radii larger than a few $M$, where the primary physical effect is compressive heating ($\rho \propto r^{-3/2}$ implies $T \propto r^{-1}$ at nonrelativistic temperatures) as gas flows in from the Bondi radius towards the center. At smaller radii ($r < M$), the gas in the four naked singularity models cools to much lower temperatures. Here the gas density is large enough that radiative cooling becomes important. Although much of the radiation is beamed towards small radii, nevertheless enough escapes to cause an enhanced luminosity at infinity. The only exception is the JMN-1 $M_0=0.7$ model where, because of the presence of a photon sphere, the amount of radiation that escapes to infinity is highly suppressed.

\subsection{Images and Shadows}

\begin{figure*}
\centering
\subfigure{\includegraphics[scale=0.87]{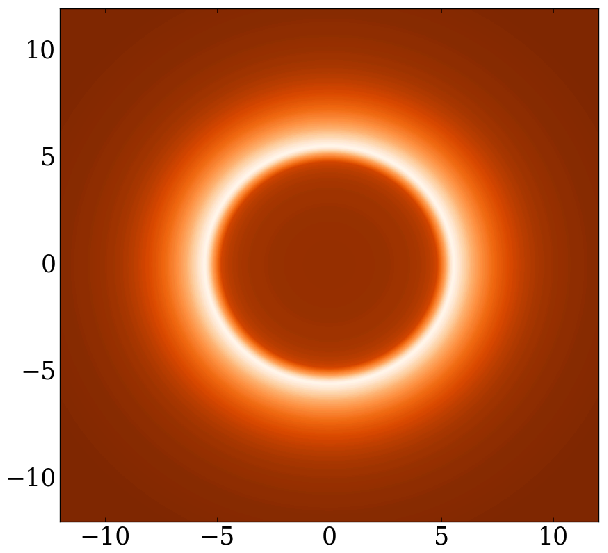}}\hspace{-1.0cm}
\subfigure{\includegraphics[scale=0.87]{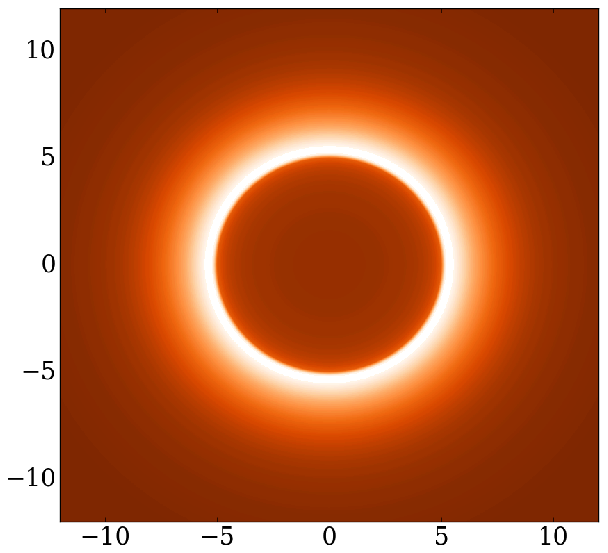}}\hspace{-1.0cm}
\subfigure{\includegraphics[scale=0.87]{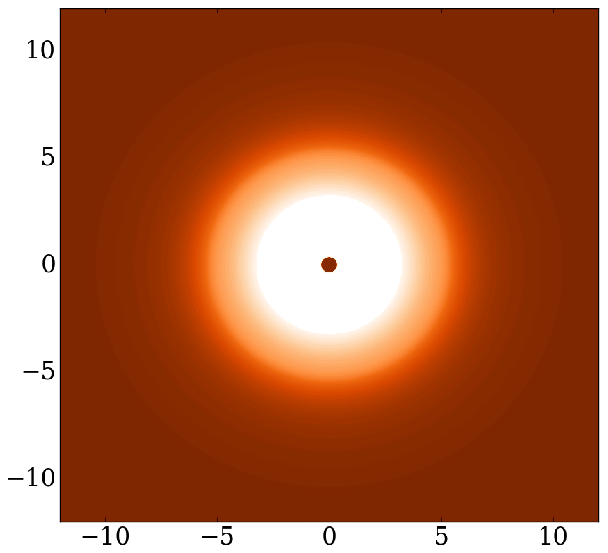}}
\subfigure{\includegraphics[scale=0.87]{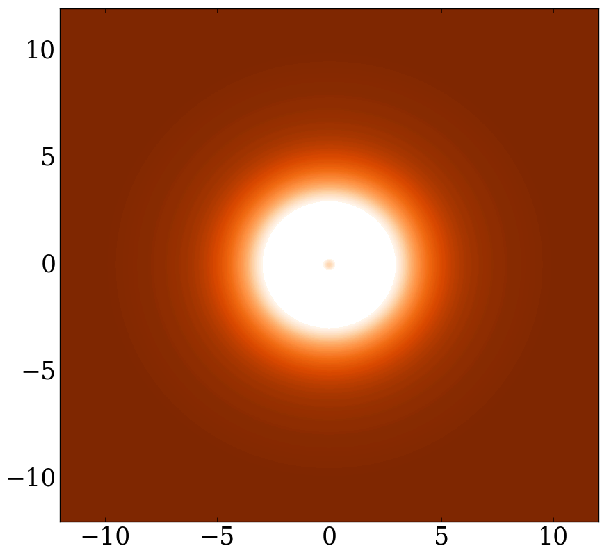}}\hspace{-1.0cm}
\subfigure{\includegraphics[scale=0.87]{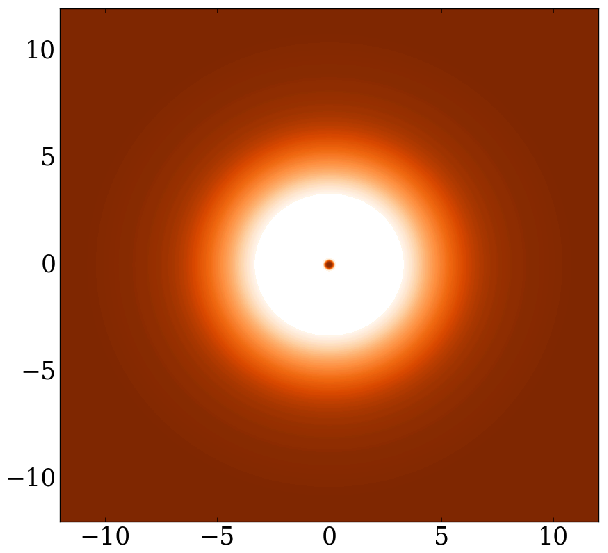}}\hspace{-1.0cm}
\subfigure{\includegraphics[scale=0.87]{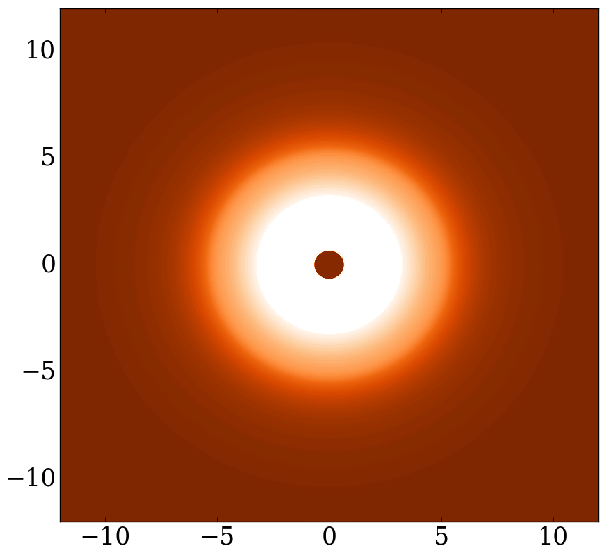}}
\caption{Shows images in the mm band (200--250\,GHz) for the accretion
  models described in \S\ref{sec:shadow2}.  All the panels use the
  same (arbitrary) color scale. Top Row: From left to right, the
  images correspond to the Schwarzschild black hole, JMN-1 naked
  singularity with $M_0=0.7$, and JMN-1 with $M_0 = 0.6$. The dark
  spot at the center of the third image is because the inner edge of
  the grid is at $r=10^{-4}M$ rather than at 0. Bottom Row: The left
  two images correspond to the JMN-2 naked singularity with
  $\lambda=0.8$ and $\lambda=0.4$, respectively. The rightmost panel
  corresponds to the same model as the one above it (JMN-1,
  $M_0=0.6$), except that the inner edge of the grid in this case is
  at $r=10^{-3}M$.}
\label{fig:images}
\end{figure*}

Figure \ref{fig:images} shows images corresponding to the accretion models under discussion. Only radiation with frequencies between 200 and 250\,GHz is considered (initial EHT results will be at
230\,GHz). The results are qualitatively similar to those shown in Figs. \ref{fig:JMN1} and \ref{fig:JMN2}.  Specifically, the Schwarzschild black hole and the JMN-1 naked singularity with
$M_0=0.7$ have well-defined dark shadows, consistent with the existence of photon spheres in these two models. The other three models, JMN-1 with $M_0=0.6$, JMN-2 with $\lambda=0.8$ and JMN-2 with $\lambda=0.4$, all have filled centers, i.e., they have ``full-moon'' images, consistent with the lack of photon spheres.

We emphasize that the accretion model considered here, which includes substantially more radiation physics, is significantly different from that in \S\ref{sec:shadow}.  Also, the images in Fig. \ref{fig:images} correspond to the mm-band, whereas in \S\ref{sec:shadow} we considered monochromatic emission and counted all the radiation. As a result, there are some quantitative differences between Fig. \ref{fig:images} and Figs. \ref{fig:JMN1}, \ref{fig:JMN2}. The rings around the shadows are somewhat narrower in the present models, and the full-moon images are somewhat smaller in angular size. Nevertheless, the qualitative results are very similar.

One feature that needs discussion is the dark spot at the center of the full-moon images in Fig. \ref{fig:images}. This is an artefact.
Because the metrics of the two JMN naked singularity
  spacetimes have power-law behavior as $r\to0$, it is necessary to
  use a logarithmic grid in $r$ when computing numerical models. As a
  result, the grid has to be truncated at some finite radius. For the
  calculations presented here, we used an inner radius of $r_{min} =
  10^{-4}M$. This is well inside the boundary radius $R_b$ where the
  naked singularity interior is matched with the Schwarzschild
  exterior.  Nevertheless, the truncation does result in a small dark
  spot at the center of the image, caused by the missing spacetime
  inside $r_{min}$.

To illustrate better the effect of this truncation, the two panels in the rightmost columnn of Fig. \ref{fig:images} show images corresponding to the same model (JMN-1 $M_0=0.6$) except that the upper panel corresponds to $r_{\rm min}=10^{-4}M$, while the lower panel corresponds to $r_{\rm min}=10^{-3}M$. The former has a smaller dark spot than the latter, confirming that the spot will disappear in the limit $r_{\rm min}\to0$.

\begin{figure*}
\centering
\subfigure[]{\includegraphics[scale=0.95]{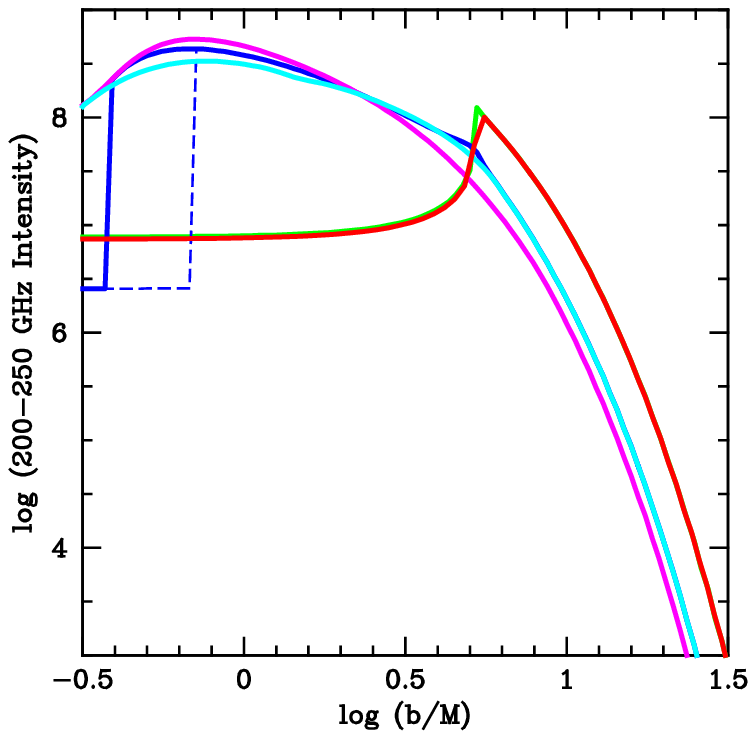}}
\subfigure[]{\includegraphics[scale=0.95]{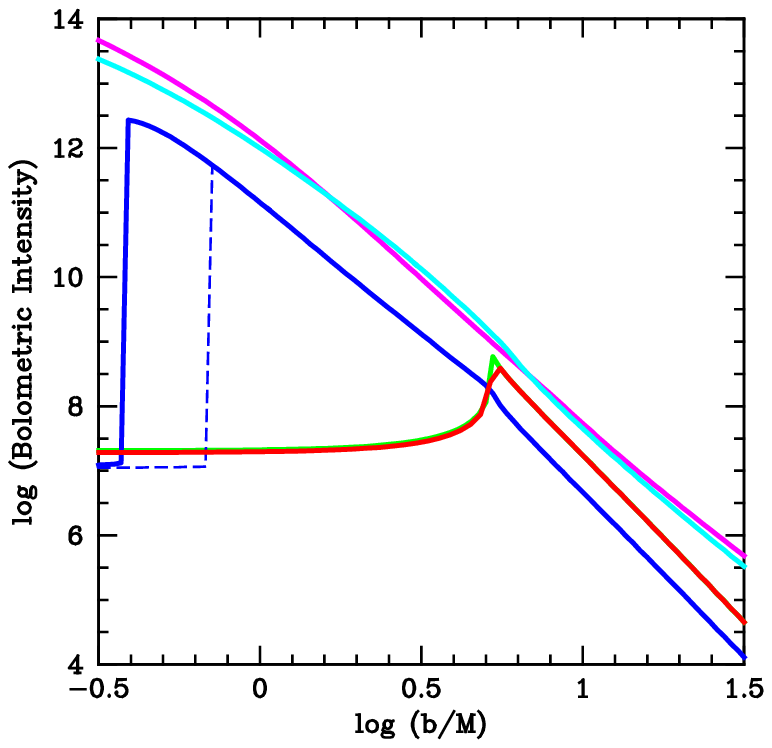}}
\vspace{-2cm}
\caption{Left: Radial profiles of the mm band (200--250\,GHz) image
  intensity versus the impact parameter $b/M$ for the Schwarzschild
  black hole (red), JMN-1 naked singularity with $M_0 = 0.7$ (green,
  under red) and $M_0=0.6$ (blue), and JMN-2 naked singularity with
  $\lambda=0.8$ (magenta) and $\lambda=0.4$ (cyan). The solid and
  dashed blue lines correspond to the same model, but with $r_{\rm
    min}=10^{-4}M$ (solid) and $r_{\rm min}=10^{-3}M$ (dashed). Right:
  Corresponding results when the bolometric radiation
  ($10^8-10^{24}$\,Hz) is considered. Note the change in the vertical
  scale.}
\label{fig:images_1D}
\end{figure*}

Figure \ref{fig:images_1D} shows radial profiles of the image intensity as a function of impact parameter for the five models. The profiles in the mm band (left panel) are quite different from those
based on the bolometric radiation (right panel). The latter are more similar to the profiles shown in Figs. \ref{fig:JMN1} and \ref{fig:JMN2} (but note that those use a linear scale whereas Fig. \ref{fig:images_1D} employs a logarithmic scale).

\section{Conclusion}
\label{sec:summary}
In this paper, we analyzed images produced by two spherically symmetric models of naked singularities, and compared them with the image produced by a spherically symmetric (Schwarzschild) black hole. We showed that naked singularities could, in some cases, cause shadows that are very similar to those produced by black holes, but in other cases, the two would have very different image structures and
would be clearly distinguishable. It follows that a careful investigation of the shadow structure will be needed before the EHT can confirm the existence of an event horizon, and thus a black hole,
in Sgr A$^*$.

To expand on the above point, even if the EHT finds a shadow in Sgr A$^*$, it will not conclusively establish the presence of a black hole in this object. The same shadow could be produced by certain naked singularity models. Among the two naked singularity models analyzed in this paper, called JMN-1 and JMN-2, we find that JMN-1 will produce shadows whenever the parameter $M_0$ (see Eq. \ref{eq:JMN1}) lies in the range $M_0\geq 2/3$. This is equivalent to the condition that the matching radius $R_b$ between the naked singularity interior spacetime and the exterior Schwarzschild spacetime satisfies $R_b<3M$, where $M$ is the mass of the object. JMN-1 models that do not satisfy the above
condition lack a shadow, and produce what we term a ``full-moon'' image. The JMN-2 model produces a full-moon image for all physically allowed choices of its parameters.

The fact that a shadow does not automatically imply an event horizon was already emphasized by \citet{Broderick-Narayan}, who showed that a model of Sgr A$^*$ with a hard surface will also produce a shadow in mm-band images. The total spectrum would, however, be different. In particular, those authors argued that observations in the infrared would easily distinguish a hard-surface model from a true black hole, because the emission from the surface would dominate in the infrared.

The naked singularity models that produce shadows, viz., JMN-1 with $M_0\geq2/3$, $R_b\leq3M$, are different in that their images and spectra at {\it all wavelengths} are nearly identical to those of a
black hole (compare the red and green curves in Figs. \ref{fig:spectra}, \ref{fig:images_1D}). Distinguishing these models will thus be much more difficult.

The full-moon image produced by the remaining naked singularity models we considered is also interesting. If such an image were observed, it would certainly rule out a black hole. Whether or not it
would confirm the presence of a naked singularity remains to be seen since other non-black hole models might also produce such images.

Similar results to those described in this paper are obtained when we consider the Schwarzschild solution with a scalar field, the so called JNW naked singularity spacetimes \citep{JNW}. For a range of parameter values, these spacetimes admit a photon sphere, and for other parameter values they do not.  In that case as well, the two kinds of models produce shadows and full-moon images, respectively. These results will be reported elsewhere.

Finally, we note that both the JMN-1 and JMN-2 models are characterized by two parameters, namely, the mass parameter $M_0$ and the matching radius $R_b$. The occurrence of a naked singularity in these models is stable with respect to variations in these two parameters, but this stability is limited since it is restricted to these specific spherically symmetric models. %
More generally, the mode stability of the JMN spacetimes as well as their stability against fluid perturbations are unproven and are currently under investigation. Because of this, the astrophysical relevance of the specific models considered here is uncertain. On the other hand, the qualitative results presented here regarding the nature of images and shadows are likely common for a wide class of naked singularity models. In this sense, however, the theoretical implications of our results are indeed astrophysically relevant. It is important to keep this in mind, since the EHT is already operational and collecting data. Specifically, we emphasize that shadows are not a consequence of event horizons, but of photon spheres. Therefore, if an object casts a shadow, it does not have to necessarily possess an event horizon.
Currently we are working on generalizing the solutions discussed here to rotating naked singularity models. It will be physically much more realistic to compare the shadow structure of such rotating naked singularities with shadows produced by a Kerr black hole. 

\bigskip\noindent
{\bf ACKNOWLEDGEMENTS}

\noindent The authors thank the International Centre for Theoretical Sciences, Bangalore, India, for hospitality during some of this work. RN was supported in part by NSF grant AST1312651, and the Black Hole Initiative at Harvard University, which is supported by a grant from the John Templeton Foundation. The authors also thank the referee for valuable comments.


\bsp	
\label{lastpage}
\end{document}